\documentclass{aa}  

\usepackage{graphicx}
\usepackage{txfonts}
\usepackage{xcolor}
\usepackage[utf8]{inputenc}
\usepackage{url}
\usepackage{amsmath}
\usepackage{commath}
\usepackage{comment}
\usepackage{hyperref}
\hypersetup{colorlinks=true,linkcolor=blue,filecolor=blue,citecolor = blue,urlcolor = blue}

\begin{document} 

    \title{Carbon envelopes around merging galaxies at z $\sim$ 4.5}

   \author{C. Di Cesare         \inst{1,2,3}\fnmsep\thanks{\email{claudia.dicesare@uniroma1.it}} \and M. Ginolfi\inst{4,5} \and L. Graziani\inst{1,2,3} \and R. Schneider\inst{1,2,3,6} \and M. Romano\inst{7,8} \and G. Popping\inst{9} 
          }

   \institute{$^1$ Dipartimento di Fisica, Sapienza, Università di Roma, Piazzale Aldo Moro 5, 00185, Roma, Italy\\ $^2$ INFN, Sezione di Roma I, Piazzale Aldo Moro 2, 00185 Roma, Italy\\ $^3$ INAF/Osservatorio Astronomico di Roma, Via di Frascati 33, 00078 Monte Porzio Catone, Italy\\ $^4$ Dipartimento di Fisica e Astronomia, Università di Firenze, Via G. Sansone 1, 50019, Sesto Fiorentino (Firenze), Italy \\ $^5$ INAF/Osservatorio Astrofisico di Arcetri, Largo E. Femi 5, 50125 Firenze, Italy \\ $^6$ Sapienza School for Advanced Studies, Viale Regina Elena 291, 00161 Roma, Italy  \\ $^7$ National Centre for Nuclear Research, ul. Pasteura 7, 02-093, Warsaw, Poland \\ $^8$ INAF - Osservatorio Astronomico di Padova, Vicolo dell'Osservatorio 5, 35122, Padova, Italy \\ $^9$ European Southern Observatory, Karl-Schwarzschild-Str. 2, D-85748 Garching, Germany}

   \date{Received: January 2024; Accepted: August 2024}

   \abstract 
   {Galaxies evolve through a dynamic exchange of material with their immediate surrounding environment, the so-called circumgalactic medium (CGM). Understanding the physics of gas flows and the nature of the CGM is thus fundamental to studying galaxy evolution, especially at $4 \leq z \leq 6$ (i.e. at post-Reionization Epoch) when galaxies rapidly assembled their masses and reached their chemical maturity. Galactic outflows are predicted to enrich the CGM with metals, although gas stripping in systems undergoing a major merger has also been suggested to play a role.}
   {In this work, we explore the metal enrichment of the medium around merging galaxies at  $z \sim 4.5$, observed by the ALMA Large Program to INvestigate [CII] at Early times (ALPINE).
   To do so, we study the nature of the [CII]158 $\mu$m emission in the CGM around these systems, using simulations to help disentangle the mechanisms contributing to the CGM metal pollution.}
   {By adopting an updated classification of major merger systems in the ALPINE survey, we select and analyze merging galaxies whose components can be spatially and/or spectrally resolved in a robust way. Thus, we can distinguish between the [CII] emission coming from the single components of the system and that coming from the system as a whole.
   We also make use of the \texttt{dustyGadget} cosmological simulation to select synthetic analogs of observed galaxies and guide the interpretation of the observational results.}
   {We find a large diffuse [CII] envelope ($\gtrsim 20 $ kpc) embedding all the merging systems, with at least $25 \%$ of the total [CII] emission coming from the medium \textit{between} the galaxies. Using predictions from \texttt{dustyGadget} we suggest that this emission has a multi-fold nature, with dynamical interactions between galaxies playing a major role in stripping the gas and enriching the medium with heavy elements.}
   {} 
   

    \maketitle
%

\section{Introduction}

The circumgalactic medium (CGM) is a buffer medium between interstellar and intergalactic media (ISM, IGM). It regulates gas flows, stellar feedback and, consequently, star formation activity and galaxy growth. Generally, one refers to the CGM as the gas outside the galaxy, but indicatively within one virial radius $R_{vir}$ of its dark matter (DM) halo. However, it is important to keep in mind that some of the processes involving the CGM, for example galactic outflows, can reach larger radii \citep{Tumlinson+2017, Faucher+Oh2023}. The CGM is a fundamental component in galaxy evolution, and studying it, both in emission and absorption provides hints on how galaxies assemble their masses through cosmic time. To understand the cycle of baryons between the ISM and the CGM we need observations of the stellar component and of the multiple gas phases both within and around galaxies (see \citet{CelineHowk2020} for a review on the baryon and metal cycles in galaxies). In this context, the Atacama Large Millimeter/submillimeter Array (ALMA) has recently opened a window to explore the cold neutral and molecular gas in early galaxies with unprecedented levels of detail \citep{Capak+2015, LeFevre+2020, Bouwens+2022}. On the other hand, we need simulations able to resolve the characteristic scales of CGM/ISM, which usually adopt zoom-in refinement schemes\footnote{Such as Adaptive Mesh Refinement \citealt{Hummels+2019,Peeples+2019} and moving mesh codes \citealt{vandeVorrt+2019}.} (see \citealt{Pallottini+2017, Lupi+2020, Lupi+2020b, Pallottini+2022}), or cosmological boxes with a moderate ($\sim 100$ pc) spatial resolution (see, for example, \citealt{Katz+2017} where the authors include on-the-fly radiative transfer and detailed non-equilibrium chemistry, to model self-consistently the ISM of high redshift galaxies). Finally, see \citet{Faucher+Oh2023} for a recent general review of the key physical processes that operate in the CGM from a theoretical point of view.

In the last decades, theories of galaxy evolution and cosmological numerical simulations have predicted that high-$z$ galaxies assemble their masses via both cold gas accretion from the IGM \citep{Dekel+2009,Topping+2022} and major mergers i.e. dynamical interactions between galaxies of nearly equal stellar masses \citep{Hopkins+2010,Schaye+2015,Duncan+2019,Romano+2021}. In particular, dynamical interactions between galaxies can drive a significant amount of gas towards the center of the interacting system boosting the efficiency of star formation - up to a factor 2 for major mergers - and also trigger starburst and active galactic nuclei (AGN) activity (\citealt{Oser+2012,LopezSanjuan2012,Kaviraj+2014,Behroozi+2015,Reeves+2023}). Moreover, ongoing and post-mergers episodes can disturb and change the morphology of the galaxies involved, leading to tails of stripped material (i.e. tidal tails), irregular shapes, and disturbed velocity fields \citep{Conselice+2003,Conselice+2008, Casteels+2014}. Tidal tails are challenging to observe at high redshift because of the diffuse and faint nature of the stripped gas. However, thanks to ALMA we can study the efficiency of gas stripping and circumgalactic gas mixing in the early Universe, by mapping the morphology and the kinematics of the CGM around merging galaxies using bright far-infrared (FIR) lines such as the singly ionized carbon (hereafter [CII]) at 158 $\mu$m, which is generally the brightest FIR emission line for star-forming galaxies \citep{CarilliWalter2013}. 

[CII] 158 $\mu$m is an important tracer of the ISM in the local and high-$z$ galaxies and, thanks to the low ionization potential (11.26 eV) of the neutral carbon, it is abundant in both the cold and warm ISM, as well as in the molecular one. Via its fine structure emission line at 158 $\mu$m ($\rm {}^2P_{3/2}^0  \rightarrow  \rm {}^2P_{1/2}^0$), it acts as a coolant in the cold ISM, then, it is considered as one of the main tracer of cold gas in galaxies, star-forming regions and molecular clouds. As a consequence, many studies focused on the investigation of the [CII] emission and, in particular, the [CII] - star formation rate (SFR) relation in the local Universe and at high-redshift, using both observations \citep{DeLooze+2014, Herrera-Camus+2015, Herrera-Camus+2018, Carniani+2018, Schaerer+2020, Romano+2022} and simulations \citep{Katz+2017, Popping+2019, Ferrara+2019, Lupi+2020, Vallini+2020, Schimek+2023}.

Observations of [CII] in early main-sequence galaxies led to the discovery of extended [CII] {\it envelopes} (up to 10 kpc) around these systems. These envelopes were at first found by stacking the [CII] emission in 18 ALMA-detected star-forming galaxies at 5 < $z$ < 7 \citep{Fujimoto+2019} and in a large sample of normal star-forming galaxies at 4 < $z$ < 6 \citep{Ginolfi+2020b}, and later confirmed, on an individual basis, by \citet{Fujimoto+2020} and \citet{Lambert+2023}. These extended emissions have been found also at higher redshifts, up to $z \sim 7 $, see for example \citet{Herrera-Camus+2021, Fudamoto+2022, Akins+2022}. However, see \citet{Posses+2023} for an example of a normal star-forming galaxy at $z \sim 7$ in which no extended [CII] has been detected.

In this paper, we present a study on the properties of [CII] emission around major merging systems at $z \sim 4.5$, with the aim of characterizing the metal enrichment of their inner CGM and investigating its dependence on the dynamical interaction between the galaxies. Our target systems are drawn from the ALMA Large Program to INvestigate [CII] at Early times (ALPINE), an ALMA large program that observed [CII] and thermal dust continuum emission for a large sample of main-sequence galaxies in the redshift range $4.4 < z < 5.9$. The interested reader can find more information on the detection strategy, the data processing, and the ancillary data of the ALPINE sample in \citet{LeFevre+2020,Bethermin+2020, Faisst+2020}, respectively. The analysis of the [CII] emission has revealed a diverse distribution of morphological and kinematical properties in the ALPINE sample \citep{Faisst+2020,Jones+2021}, with the detection of signatures of metal-enriched gas outflows powered by star formation activity in the high-velocity tails of the stacked [CII] emission spectrum \citep{Ginolfi+2020b}. Also, a detailed morpho-spectral decomposition analysis in one of the ALPINE systems undergoing major merging has revealed the presence of a widespread [CII] emission component, extending to scales of a few tens of kpc, which has been interpreted as a possible signature of processed ISM stripped by the strong gravitational interaction, suggesting that mergers could be an efficient mechanism of metal enrichment and gas mixing in the CGM around high-$z$ galaxies \citep{Ginolfi+2020a, Jones+2020}. 

Motivated by these findings, we apply a similar morpho-spectral decomposition analysis on additional systems classified as mergers in the ALPINE sample. To this aim, we make use of the merger classification of ALPINE galaxies carried out by \citet{Romano+2021}. 
The observational part of this work is complemented with predictions from cosmological simulations run with the hydrodynamical code \texttt{dustyGadget} \citep{Graziani+2020}, which guide our interpretation of the results. We would like to clarify that throughout the paper when presenting results based on simulations, we use the mass of carbon in the cold ISM medium (T $\lesssim 10^4$ K) as a proxy to estimate the expected [CII] emission. Thus, we do not employ any radiative transfer code to predict the [CII] luminosity, and we do not assume any empirical relation to link the physical properties of galaxies with the expected [CII] luminosity.

The paper is organized as follows: in Section \ref{sec:observations} we describe the selected observational sample and the analysis we carried out; Section \ref{sec:diffEmission} presents the [CII] diffuse emission that we find in the observed candidates, and its interpretation; in Section \ref{sec:simulations} we introduce the hydrodynamical code \texttt{dustyGadget}, explain the adopted procedure to identify synthetic major mergers, their time evolution, and their gas distribution. Finally, we exploit the predictions from \texttt{dustyGadget} to interpret the observed [CII] diffuse emission, in Section \ref{sec:discussion}. Our conclusions are drawn in Section \ref{sec:conclusion}.

Throughout the paper, we assume a flat $\Lambda$CDM cosmology with cosmological parameters from \citet{PLANK+2016} consistent with that assumed by \texttt{dustyGadget} simulation (see Section \ref{sec:simulations}) and adopt a \cite{Salpeter+1955} initial mass function (IMF). All the stellar masses and star formation rates in this paper have been converted to a \citet{Salpeter+1955} IMF following the conversion factors from \citet{Madau&Dickinson2014}. At $z \sim 4.6$ - that is the mean redshift of our sample - 1 arcsecond corresponds to 6.69 proper kpc.

\section{Observational sample and data processing}
\label{sec:observations}
In this section we first briefly introduce the ALPINE survey and the properties of the target galaxy sample; and then discuss the classification done by \citet{Romano+2021} on merging systems, and how we selected - among these - major mergers needed for the aim of this work. Finally, we describe the ALMA data reduction procedure and the analysis we performed on our sample of major merging galaxies. 

\begin{table*}
    \renewcommand{\arraystretch}{1.5}
    \centering
    \caption{Physical parameters of the major merging galaxies selected for this work. Together with the source ID and redshift estimates for both the components of the merger ($z_1$ and $z_2$), we also include the velocity offset ($\Delta v $) and projected distance ($r_{\rm p}$) between the merger components and the [CII] flux ratio ($\mu _{\rm [CII]}$) and $\rm K_s$ band flux ratio ($\mu _{\rm K}$) \citep{Romano+2021}. The last two columns include the estimates for stellar masses and star formation rates for the entire merging system as estimated by \citet{Faisst+2020} (see their paper for details on the photometry and SED fitting procedure), once we converted the IMF from \citet{Chabrier2003} to \citet{Salpeter+1955}.}
    \begin{tabular}{c|c|c|c|c|c|c|c|c}
        \hline
        source ID & $z_1$ & $z_2$ & $\Delta v \: [\rm km/s]$ & $ r_p \: [\rm kpc]$ & $\mu _{\rm [CII]}$ & $\mu _{\rm K}$ & Log($\rm M_\star / M_\odot$) & Log($\rm SFR / M_\odot yr^{-1}  $) \\       
        \hline
        \hline
        DC\_818760 & 4.5626 & 4.5609 & 92.3 & 9.9 & 1.3 & 2.6 & 10.85$^{+0.11}_{-0.10}$ & 2.88$^{+0.19}_{-0.25}$ \\
        DC\_873321 & 5.1545 & 5.1544 & 4.5 & 6.5 & 1.2 & 3.1 & 10.18$ ^{+0.13}_{-0.16}$ & 2.16$^{+0.22}_{-0.17}$ \\
        vc\_5100541407 & 4.5628 & 4.5628 & 1.9 & 13.8 & 1.6 & 1.4 & 10.33$^{+0.14}_{-0.15}$& 1.74$^{+0.26}_{-0.23}$\\
        vc\_5100822662 & 4.5210 & 4.5205 & 22.3 & 10.9 & 1.6 & 1.7 &10.39$^{+0.13}_{-0.14}$& 2.02$^{+0.23}_{-0.24}$ \\
        vc\_5101209780 & 4.5724 & 4.5684 & 217.3 & 10.8 & 4.1 & 2.5 &10.27$ ^{+0.12}_{-0.12}$& 1.79$^{+0.25}_{-0.21}$ \\
        vc\_5180966608 & 4.5294 & 4.5293 & 8.9 & 7.2 & 3.0 & 3.7 &11.04$^{+0.12}_{-0.13}$& 2.35$^{+0.27}_{-0.25}$ \\
        \hline
    \end{tabular}
    \label{tab:OBStargets}
\end{table*}

\subsection{Targets selection}
\label{targets}
The ALPINE survey is designed to detect the [CII] line at 158 $\mu$m rest-frame and the surrounding FIR continuum emission for a sample of 118 normal galaxies at $4.4 < z < 5.9$. Its targets are selected from the Cosmic Evolution Survey (COSMOS; \citealt{Scoville+2007a,Scoville+2007b}) and Extended Chandra Deep Field South (E-CDFS; \citealt{Giavalisco+2004,Cardamone+2010}) fields. In the following, we use "vc" and "DC" to respectively refer to vuds\_cosmos and DEIMOS\_COSMOS sources. Since these fields have been targeted by several observational campaigns, a wealth of ancillary multi-wavelength photometric data (from rest-frame UV to FIR) is available, which made it possible to recover physical properties such as stellar masses ($\rm M_\star$) and SFR through spectral energy distribution (SED)-fitting; these estimates have been performed adopting a Chabrier IMF (\citealt{Chabrier2003}, see \citealt{Faisst+2020} for a detailed description). With stellar masses in the range $9 \leq \rm Log(M_\star/M_\odot) \leq 11$ and star formation rates of $1 \leq \rm Log(SFR/M_\odot yr^{-1}) \leq 3$, 
ALPINE galaxies lie on the so-called main sequence of star-forming galaxies \citep{Rodighiero+2011,Tasca+2015}, being thus representative of the underlying galaxy population at $\rm z \sim 5$ \citep{Speagle+2014}.

To study the metal enrichment of the CGM around high-$z$ merging galaxies we exploit the work of \citet{Romano+2021}, who combined the morpho-kinematic information provided by the [CII] emission with archival multi-wavelength photometry to identify merging systems in the ALPINE sample, and the fraction of major mergers within this subset (see also \citealt{LeFevre+2020,Jones+2021}). According to their classification criteria, \citet{Romano+2021} found that the ALPINE sample is composed of 31\% of mergers (23 out of 75 \footnote{Among the 118 main sequence galaxies of the ALPINE sample, 75 are detected in [CII] at S/N > 3.5 \citep{Bethermin+2020}.}), leading to a major merger fraction of $f_{\rm MM} \sim 0.44$ (0.34) at $ z \sim 4.5$ (5.5). This result is in good agreement with morphological studies by \citet{Conselice+2009} at the same redshift and, when combined with other works down to the Local Universe, suggests a rapid increase in the cosmic merger fraction from $z = 0$ to $ z \sim 2$, a peak at $ z \sim 2-3$ and a possible slow decline for $ z \geq 3$. 

For the aim of this study, starting from the classification by \citet{Romano+2021}, we selected major merging systems that are spatially and/or spectrally separated in a robust way. In particular, we looked for merging systems with: 
\begin{enumerate}
    \item a velocity separation $\Delta v \leq 500 \: \rm km/s$, so that the two systems can be considered as gravitationally bound (see \citealt{Patton+2000,Lin+2008,Ventou+2017});
    \item a projected distance $r_{\rm p} > 4 \: $kpc, where $r_{\rm p} = \theta \times d_{\rm A}(z_m)$, $\theta$ is the angular separation in arcsec in the sky between the two galaxies, and $d_{\rm A}(z_m)$ is the angular diameter distance (in $\rm kpc \: arcsec^{-1}$) calculated at the mean redshift $z_{\rm m}$ of the two sources (see \citealt{Romano+2021} for more details). With this condition, the projected distance between the merging systems is larger than the typical [CII] size of individual galaxies, which on average is estimated to be $\sim 2.1$ kpc for ALPINE galaxies at these redshifts \citep{Fujimoto+2020, Romano+2021}. Indeed, closer components could just be clumps of star formation within the same galaxy, affecting the morphology and kinematics of [CII] emission;
    \item a relative stellar mass ratio of $\rm 1 < \mu _{\rm K} < 4$. Here $\mu _{\rm K}$ is defined as the ratio between the UltraVista $\rm K_s$-band fluxes of the merging components\footnote{In general, $\mu$ can be defined as :
\begin{equation}
    \mu = X^{i}_1 / X^{i}_2
\end{equation}
where $X^{i}_1 $ and $ X^{i}_2$ are the physical properties ($X^i$ = stellar mass; [CII] fluxes; $\rm K_s$-band fluxes) of the primary and secondary galaxy, with $X^{i}_1 > X^{i}_2$.}, which is used as a proxy for the mass ratio of galaxies (hereafter, $\mu _K \equiv \mu _\star$) since the $\rm K_s$-band flux is a good tracer of the stellar mass of galaxies up to $\rm z \sim 4$ \citep{Laigle+2016}.
\end{enumerate}

By applying these criteria to the 23 merging systems selected by \citet{Romano+2021}, we end up with a sample of six targets. We shall note that the $\rm K_s$-band ratio is available for 9 out of 23 merging systems; for the other 14 systems, only the [CII] flux ratio, $\mu_{\rm [CII]}$, is available, from which \citet{Romano+2021} cannot draw conclusions about the nature of the merger (see their Section 4 for more details).
Table \ref{tab:OBStargets} lists the 6 observational targets selected for our study, and summarizes their merging properties ($\Delta v$, $ r_{\rm p}$, $\mu_{\rm [CII]}$, $ \mu_{\rm K}$; \citealt{Romano+2021}) and their M$_\star$ and SFR as estimated by \citet{Faisst+2020}. 

\begin{figure*}[!t]
\centering
\includegraphics[scale=0.4]{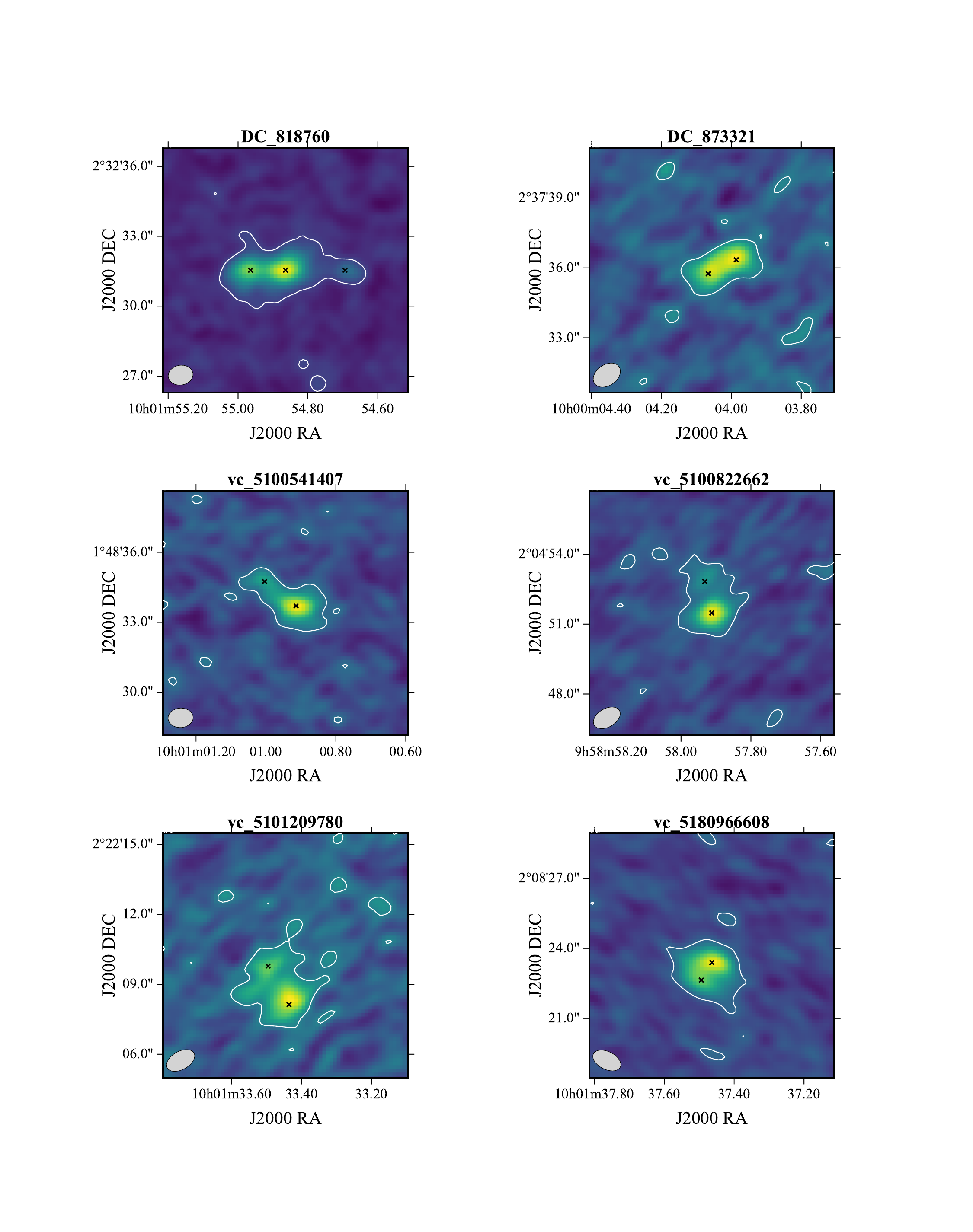}
\caption{Total velocity integrated [CII] map of the six selected systems. On top of each panel is the ID of the source. White contours indicate the positive significant level at 2$\sigma$ of [CII] emission. Black crosses mark the center of each galaxy (see the text for more information). The ALMA beam size is given in the bottom-left corners.}
\label{fig:mom0_maps}
\end{figure*}
\begin{figure*}[!t]
\centering
\includegraphics[scale=0.35]{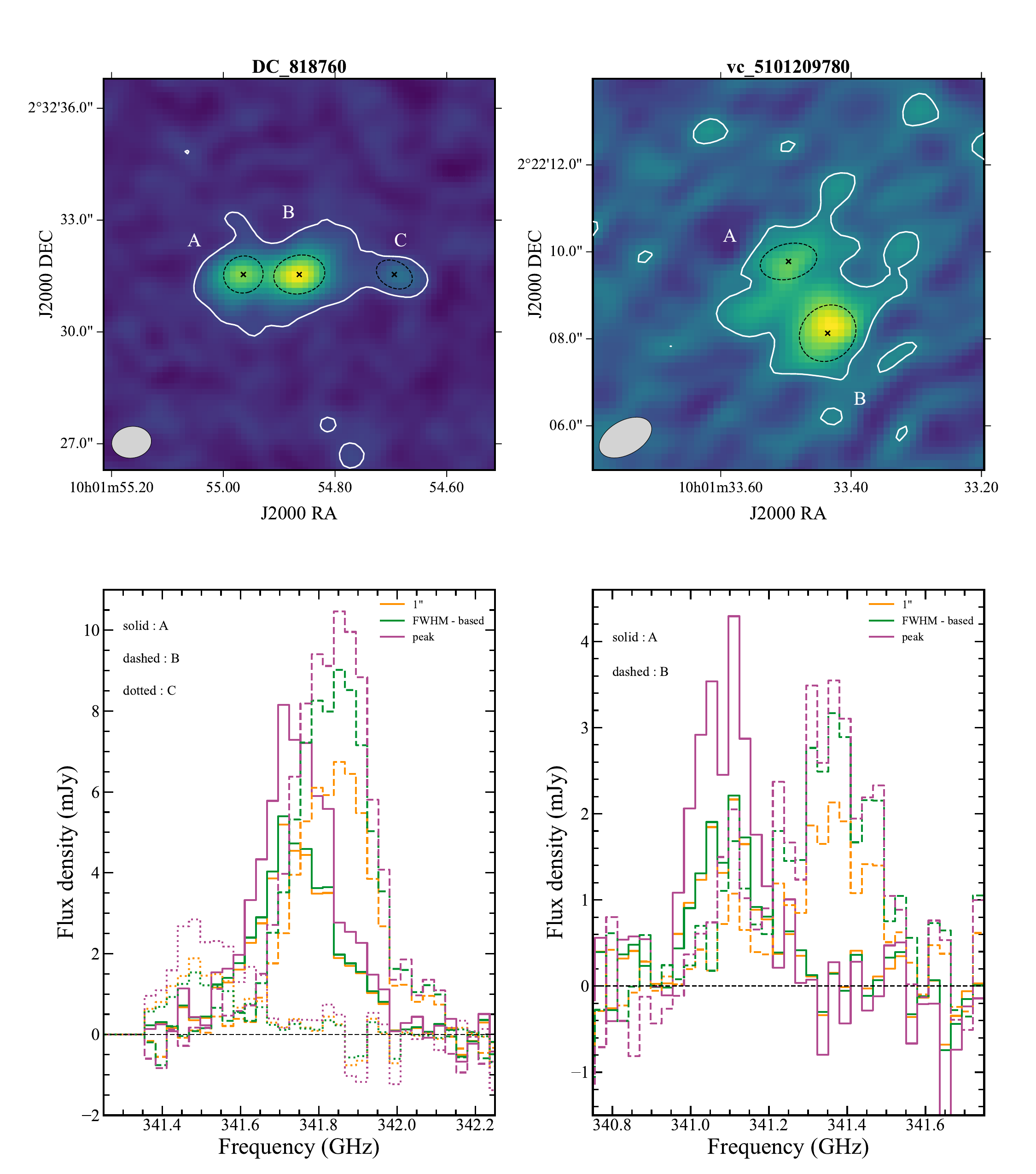}
\caption{\textit{Upper panel:} Same as Figure \ref{fig:mom0_maps}, but for two specific systems: (a) DC\_818760 (left) and (b) vc\_5101209780 (right). The white contour indicates the positive significant level at 2$\sigma$ of [CII] flux, where $\sigma _{\rm [CII]} = 84 \: \rm mJy \: km/s$ in (a) and $\sigma _{\rm [CII]} = 95 \: \rm mJy \: km/s$ in (b). The dashed black ellipses indicate the $\rm FWHM_x \times FWHM_y $ regions obtained by 2D Gaussian models and correspond to the apertures used to extract the [CII] spectra of (a) and (b); finally, the black crosses mark the center of each ellipse (i.e. the center of each galaxy). The ALMA beam size is given in the bottom-left corners. \textit{Lower panel:} [CII] spectra for each component of the merging system (different line styles) extracted using 1" (orange), FWHM-based (green), and peak apertures (pink).}
\label{fig:example}
\end{figure*}

\begin{figure*}[!t]
\centering
\includegraphics[scale=0.35]{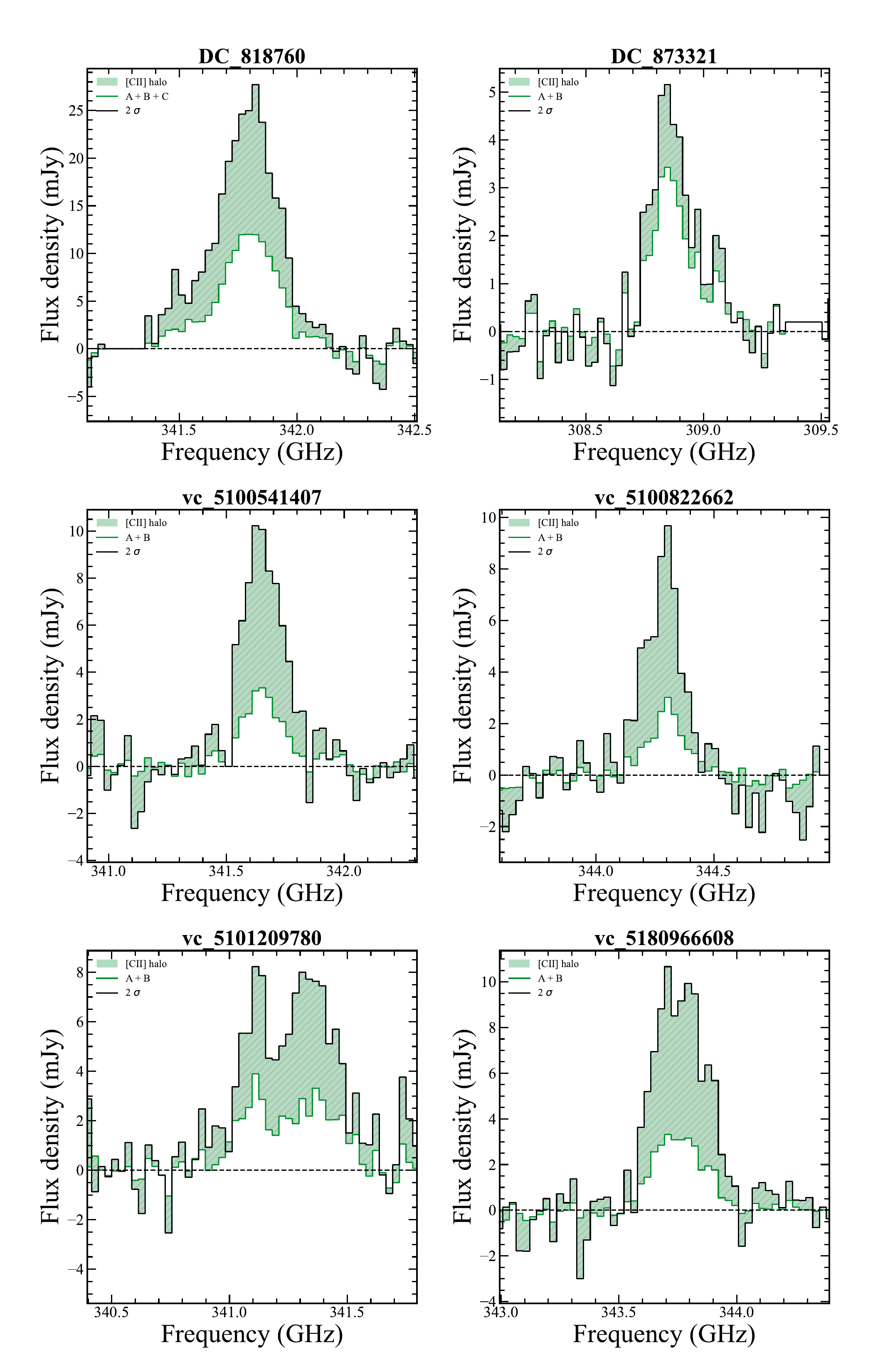}
\caption{The [CII] flux density in mJy as a function of frequency (GHz) for the galaxies in our sample. Black lines are the total [CII] emission arising from the full system (galaxy + diffuse [CII] halo) and green solid lines are the sum of the emissions coming from each component of the merging system when we consider the $\rm FWHM_x \times FWHM_y$ aperture. Green hatched areas show the emission coming from the diffuse [CII] halo and the [CII] fraction is in green boxes  (see Section \ref{sec:diffEmission} for the definition of this quantity and discussion).}
\label{fig:spectra}
\end{figure*}

\subsection{Observational analysis}
\label{sec:OBSanalysis}
In this section, we describe processing and analysis performed on the major merging systems whose properties have been reported in Table \ref{tab:OBStargets}. 

Firstly, we reduce the ALMA data for these targets using the Common Astronomy Software Applications (CASA; \citealt{McMullin+2007}) pipeline. Each data cube is continuum-subtracted using the CASA task \texttt{uvcontsub} over the line-free visibilities in all spectral windows to obtain line-only cubes. The [CII] datacubes are generated from the continuum-subtracted visibilities using the task \texttt{tclean} until we reach a S/N < 2 for the residuals. We choose a natural weighting of the visibilities to maximize the sensitivity, a common pixel size of 0.15", and a common spectral bin of $25 \, \rm km \: s^{-1}$ (beam size of $\sim 1 "$).  

Secondly, we visually inspect ALMA datacubes, looking for [CII] emissions coming from the components of each merging system. Every time we find such an emission, we adopt a $1"$ aperture centered on the emitting galaxy to extract a preliminary [CII] spectrum, which is then fitted using a single 1D Gaussian model.  

Finally, we consider the $2\sigma$ confidence interval of the Gaussian fit to get the min-max frequency range that we collapse to generate the moment-0 (i.e. velocity integrated) maps of each merging component (dubbed as \#A, \#B, etc) using the CASA \texttt{immoments} task. The moment-0 map of the entire system is then obtained by collapsing the absolute minimum and maximum of the previously obtained frequency (i.e. $min [\rm min_{freq}^A,\rm min_{freq}^B]$ and $max[\rm max_{freq}^A,\rm max_{freq}^B]$). 

Figure \ref{fig:mom0_maps} shows the total velocity-integrated [CII] maps for our sample; white contours indicate the 2$\sigma$ region\footnote{The 2$\sigma$ region is computed as the standard deviation in the total moment-0 map once we mask the source.} and black crosses are the centers of each emitting component (see procedure in the following), and all the S/N are consistent with those of the ALPINE survey \citep{Bethermin+2020}. Having the moment-0 maps we then fit a single 2D Gaussian model to each merging component, masking it, and retrieving morphological information such as the coordinates of the [CII] emission peak, the full-width half maximum (FWHM) of the major and minor axis of the Gaussian and its position angle (PA). For one system, DC\_873321, we had to perform a two 2D Gaussian components fit, because of the little spatial separation and similar [CII] luminosity ($ \mu_{[ \rm CII]} \sim 1$; see Table \ref{tab:OBStargets}) of the merging galaxies, which prevent us from fitting single 2D Gaussian models. Once we estimate the coordinates of the [CII] emission peaks of each source, we use it to center different size apertures. In particular, we employ a $1"$ aperture, a 2x2 pixel square aperture \footnote{In this case, the flux is the mean Jy/beam for 2x2 pixels region.} (hereafter "\textit{peak}"), and $\rm FWHM_x \times \rm FWHM_y$ aperture (hereafter "\textit{FWHM-based}") - with x and y being respectively the major and minor axis of the 2D Gaussian model. From these apertures, we extracted the [CII] fluxes in $ mJy \; km / s$. In this analysis, we assume that all the [CII] emission coming from the apertures is associated with the merging galaxies, while that coming from the 2$\sigma$ contour is the total emission (galaxies + diffuse [CII] emission). 

Figure \ref{fig:example} shows, in the upper panels, the total velocity-integrated [CII] maps for two systems highlighting the FWHM-based apertures for each merging component (dashed black lines), the center of each galaxy (black crosses) and the 2$\sigma$ regions (white contours) and, in the lower panels, the [CII] flux density. In particular, different line styles correspond to each merging component, while different colors (orange, green, and pink) to line spectra extracted considering $1"$, \textit{FWHM-based} and \textit{peak} apertures respectively. Looking at the spectra we note that the brightest emission is the one coming from the \textit{peak} aperture (pink) i.e. the center of each galaxy, while $1"$ and \textit{FWHM-based} apertures in some cases have comparable emissions (see for example component \#A of both DC\_818760 and vc\_5101209780). 

The merging system DC\_818760 is the only triple merger in our sample. The bottom left panel of Figure \ref{fig:example} shows that the emission coming from component \#C is $\sim$ 4 and 5 times dimmer than the one coming from components \#A and \#B respectively. The major merger is indeed happening between galaxies \#A and \#B, which are closely associated both spatially and in velocity, while \#C is identified as an upcoming minor merger (see \citealt{Jones+2020} for a detailed discussion about this system).

Finally, Figure \ref{fig:spectra} shows the [CII] emission arising from the total system (black) together with the sum of the emissions arising from each component of the merging system when we adopt the \textit{FWHM-based} aperture (green). The difference between these two emissions (green hatched area) can be interpreted as due to the diffuse [CII] envelope around the galaxies. Already in \citet{Ginolfi+2020a} the authors analyzed in detail the merging system vc\_5101209780, finding that about 50\% of the total [CII] emission arises from a gaseous envelope distributed between the individual components of the system. 

From the analysis of the observational sample used in this work, we find that a consistent fraction of the total [CII] emission arises from the diffuse gas \textit{between} the galaxies, as qualitatively shown in Figure \ref{fig:spectra}.

DC\_873321 is the system with the least [CII] emission associated with the diffuse component, according to Figure \ref{fig:spectra} (top right panel), and it is also the only system where we had to perform a two components 2D Gaussian fit instead of two \textit{single} component 2D Gaussian models (see previous discussion). Because of the difficulty in separating the emission from the galaxies and that of the surrounding medium, it is likely that part of the [CII] emission from the diffuse halo is associated with the individual components of the system. Note that, systems in which it is more difficult to separate the components may be in a more advanced phase of the merger (i.e. a closer interaction).

In Appendix \ref{app:vel_shift} we test the presence, if any, of a velocity shift among the diffuse and galactic components of the [CII] emission.

\section{[CII] emission from the CGM}
\label{sec:diffEmission}
In this section, we investigate the amount of [CII] emission arising from the diffuse halo in a more quantitative way, in order to identify possible trends between the diffuse emission and the physical properties of the merging galaxies, such as their [CII] luminosity ($\rm L_{[CII]}$), stellar mass (M$_\star$) and SFR. We also explore possible trends between the fraction of [CII] emission associated with the diffuse component, or inner CGM of the galaxies, and the relative properties of the merging systems, such as $r_{\rm p}$, $\rm \mu_{\star}$ and $\rm \mu_{[CII]}$. Indeed, these trends can give us hints about the nature of the diffuse [CII] envelope we observe, helping us to understand if this originates from metal-enriched gas outflows, from tidally stripped material during the gravitational interaction between the merging galaxies, and/or it pertains to small satellites lying around the merging galaxies (see discussion in \citealt{Ginolfi+2020a}). 

\begin{figure*}
\centering
\includegraphics[scale=0.3]{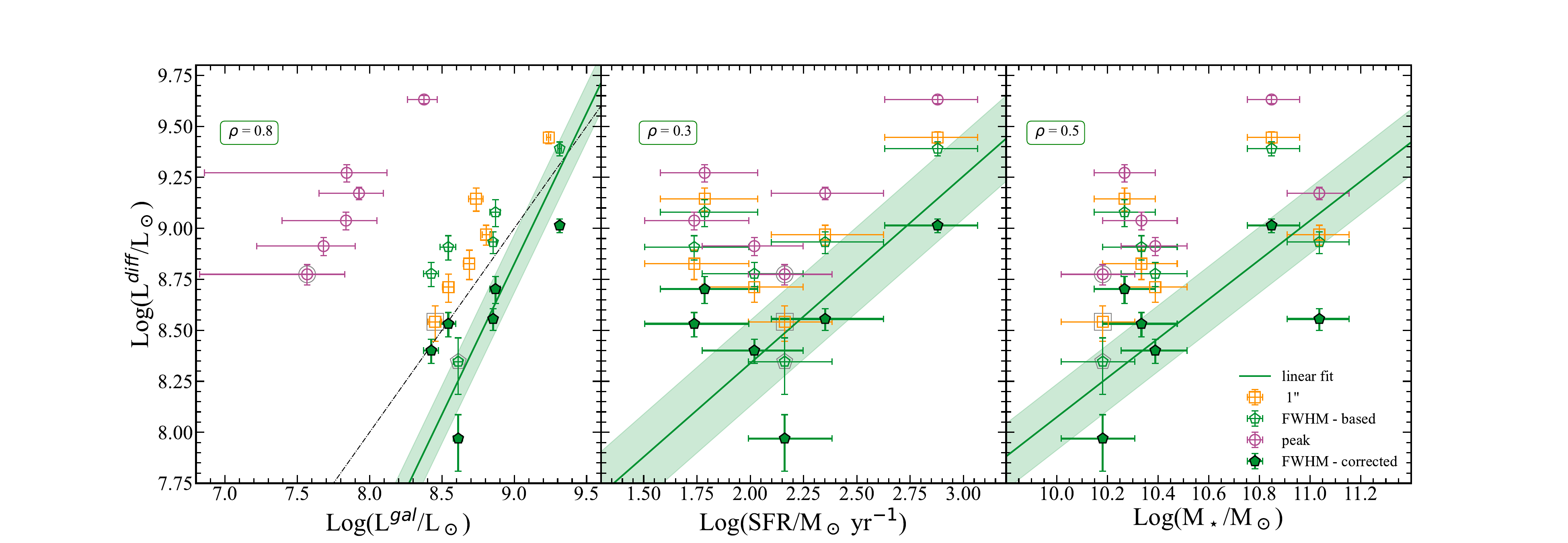}
 \caption{Trends between the [CII] luminosity coming from the diffuse halo around merging systems, L$^{\rm diff}$, and their integrated physical properties. Different colors refer to different apertures adopted to extract the emission from the merging galaxies, L$^{\rm diff, i}$, with i = 1$"$ (orange border), FWHM-based (green border), and peak (pink border). In each panel, the filled green pentagons are the estimates obtained after applying the correction for the tails of the ALMA PSF (see text for discussion); the green solid line and shaded areas are respectively the linear fit and the associated error for these estimates. We also report the Spearman coefficient ($\rho$) of the analyzed relation. \textit{Left:} L$^{\rm diff}$ is shown as a function of the [CII] luminosity coming from the galaxies, L$^{gal}$. The black dot-dashed line is the 1:1 relation. \textit{Middle and Right:} L$^{\rm diff}$ as a function of the total SFR and total M$_\star$ of the merging systems (see Table \ref{tab:OBStargets}). Highlighted in gray is DC\_873321, see the main text for the discussion on this candidate.} 
\label{fig:diffuseCGM}
\end{figure*}
In order to estimate the \textit{total} [CII] fluxes \- in $ \rm Jy \: km/s$ \- we extract the emission from the 2$\sigma$ region and fit it using a single Gaussian model. We then convert the measured fluxes in [CII] luminosity following the relation \citep{Solomon+1992,CarilliWalter2013}:

\begin{equation}
\label{eq:lumFreq}
{\rm L_{[CII]}} = 1.04 \times 10^{-3} \; {\rm F_{[CII]}} \; D_{L}^2(z) \; \nu _{obs}(z)
\end{equation}
\noindent
where $\rm L_{\rm [CII]}$ is in $ \rm L_\odot$, $D_{L}(z)$ is the luminosity distance (in Mpc) at the redshift of the merging system and $\nu _{obs}(z)$ is the observed frequency of the [CII] line (in GHz). We dub the total luminosity of the merging system $\rm L^{2\sigma}_{[CII]}$, and we estimate the error associated with this luminosity as $ \rm \sqrt N \times rms$ of the moment-0 map, where $N$ is the number of ALMA beams enclosed in that region. To quantify the emission associated with the individual merging galaxies, we extract the spectra of each component using three different apertures centered on each galaxy (see Section \ref{sec:OBSanalysis} for a description of the adopted method). For a fixed aperture, we sum together the spectrum of each component and integrate it into the 2$\sigma$ confidence interval found before. Using Equation \ref{eq:lumFreq}, we obtain the luminosity from the galaxies for each of the three apertures (1", FWHM-based, peak). 
We use the formalism $\rm L^{i}_{\rm [CII]}$, with i = \{1$"$, FWHM-based, peak\}. The error associated with the flux coming from these apertures is consistent with the rms of the moment-0 map since the apertures are comparable to the ALMA beam.

\begin{figure*}
\centering
\includegraphics[scale=0.3]{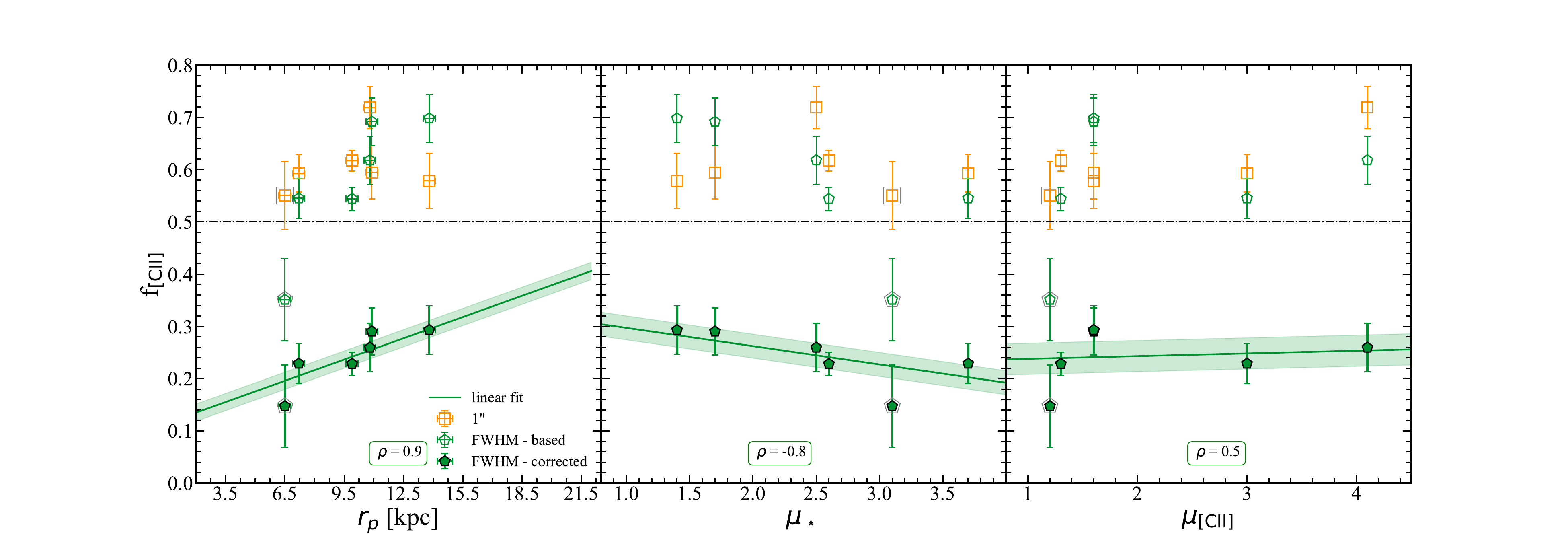}
 \caption{Trends between the fraction of [CII] emission coming from the inner CGM ($f_{\rm [CII]}$, see Equation \ref{eq:fcgm}) and the properties of the merging systems, such as $r_{\rm p}$ between the galaxies (\textit{Left}), the stellar mass ratio ($\mu_\star$, \textit{Middle}), and the [CII] emission ratio ($\mu_{\rm [CII]}$, \textit{Right}). We show the results obtained when adopting different apertures ($1"$ orange border, and FWHM-based green border) needed to define the [CII] emission associated with each component of the merger. Filled green pentagons are the estimates obtained after applying the correction for the tails of the ALMA PSF (see Section \ref{sec:diffEmission} for a discussion). In each panel, the green solid line and shaded areas are respectively the linear fit and the associated error for the more conservative estimates,  and we specify the Spearman coefficient ($\rho$) of the corresponding relation. The horizontal dot-dashed black line indicates the 0.5 value and the system DC\_873321 is highlighted in gray (see text).} 
\label{fig:CGMfraction}
\end{figure*}

Having estimated the total [CII] luminosity of the merging system and the luminosity coming from the galaxies, we compute the [CII] emission coming from the diffuse medium as:
\begin{equation}
\rm    L^{\rm diff,i} \equiv L^{2\sigma} - L^{i}
\end{equation}
\noindent
where we have dropped the subscript [CII]. We also define the fraction of [CII] emission coming from the inner CGM of the merging system (hereafter, $\rm f_{[CII]}$) as:
\begin{equation}
\rm    f_{\rm [CII]}^i \equiv \frac{L^{\rm diff,i}}{L^{2\sigma}}.
    \label{eq:fcgm}
\end{equation}
\noindent
Figure \ref{fig:diffuseCGM} shows the relation between $\rm L^{\rm diff}$ and some integrated properties of the system, such as the total [CII] luminosity associated with galaxies, the star formation rate, and the stellar mass of the entire merging system listed in Table \ref{tab:OBStargets} (see \citealt{Faisst+2020} for the discussion on photometry and SED fitting procedure). For each of these relations, we compute the Spearman coefficient, $\rho$, to see how reliable the suggested trend is: if $\rho \sim 1$ a strong correlation between the quantities is present, while for $\rho \sim -1$ there is a strong anti-correlation. Although we report the results obtained for each of the three apertures, we primarily base our conclusions on the more appropriate FWHM-based method. However, this method needs to be revised to account for possible contamination of the ALMA beam, as discussed below.
In detail, we apply a correction to address the potential flux loss (associated with individual galaxies) that might be located in the tails of the ALMA beam during the 2D Gaussian fit on the velocity-integrated maps to define the FWHM-based aperture (see Figures \ref{fig:example} and \ref{fig:spectra}). This effect could enhance the [CII] emission linked to the diffuse envelope.
To address this, we simulate artificial velocity-integrated maps of point-like sources by injecting the ALMA beams and the measured noise of the ALPINE observations, thereby matching the depth of the observed maps and replicating the same peak S/N and separations as in the observations. We then perform the same analysis on these simulated maps as we do for the FWHM-based case in the observations. Our findings indicate that the tails of the PSF indeed enhance the [CII] emission from the diffuse medium.
Consequently, our FWHM-based results need to be corrected by a factor of 0.58 (i.e., on average, about 58\% of the extended emission can be attributed to the tails of the PSF). The revised values are shown as filled green pentagons in Figures \ref{fig:diffuseCGM} and \ref{fig:CGMfraction}. We note that this results in a conservative correction as it is based on the moment-0 maps of the entire emitting regions (Figure \ref{fig:mom0_maps}) which are not optimized for the extended emission.

Figure \ref{fig:diffuseCGM} shows that there is a strong positive correlation ($\rho = 0.8$) between the [CII] luminosity of the diffuse medium and that of the merging components. This suggests that the objects with higher L$_{\rm [CII]}$ result in a more carbon-polluted medium (i.e. higher values of diffuse L$_{\rm [CII]}$). This is not unexpected, as more massive (luminous) galaxies are more star-forming and gas-rich, resulting in stronger outflows \citep{Ginolfi+2020a}. Moreover, as most of the points in the L$^{\rm diff}$-L$^{ \rm gal}$ plane, place themselves above the 1:1 relation (black dot-dashed line) we can conclude that most of the [CII] emission is associated with the diffuse medium. 

We also find a tentative ($\rho = 0.3$) relation between the diffuse [CII] luminosity and the SFR of the system, and a stronger positive correlation ($\rho = 0.5$) between $\rm L^{ diff}$ and the stellar mass of the system (see the middle and right panels of Figure \ref{fig:diffuseCGM}). These results provide some tentative indications of the origin of the metal-enriched gas powering the diffuse [CII] luminosity. In fact, stronger gas outflows are expected in more star-forming systems \citep{Ginolfi+2020a}, while more massive galaxies trace the possible presence of small unresolved satellite galaxies and/or a stronger dynamical interaction between the merging components and are associated with more star forming (SF) galaxies resulting in a greater outflows activity. The tentative trends found so far suggest that the inner CGM is polluted with heavy elements by means of outflows, dynamical interactions, and small star-forming satellites. In particular, stripping mechanisms seem to have an important role in this scenario as we observe more extended [CII] envelopes in interacting systems compared to non-interacting galaxies with similar SFR - thus with comparable outflow activity (see for example [CII] size estimates in \citealt{Fujimoto+2020}). This is also supported by the relation between f$_{\rm [CII]}$ and $\mu _\star$ (see Figure \ref{fig:CGMfraction} and its discussion) which suggests that systems with the strongest gravitational interaction, i.e. $\mu _\star$ closer to unity, have a higher fraction of [CII] emission coming from the diffuse envelope.

In Figure \ref{fig:CGMfraction} we further explore the dependence of the fraction of [CII] emission coming from the inner CGM of the merging systems, $\rm f_{[CII]}$, and the properties of the merging pairs, such as $r_{\rm p}$ between the galaxies, $\mu _\star$, and $\mu _{\rm [CII]}$. In each panel, the green solid line and shaded areas are respectively the linear fit and the associated error when considering the ALMA PSF tails contribution to the diffuse emission. In Figure \ref{fig:CGMfraction} we do not include the peak aperture as we obtain $\rm f_{[CII]} \sim 0.95$ in all the systems given that, in this case (i.e. unresolved merging galaxies), most of the [CII] emission is associated with the inner CGM.

From this figure, we note that the fraction of [CII] emission coming from the inner CGM of the merging systems is in the range $\rm f_{[CII]} \simeq 0.14 - 0.30$ ($\sim 25 \%$, on average) when accounting for the possible effect of the PSF broadening as discussed in Section \ref{sec:diffEmission}. The black horizontal line corresponds to $\rm f_{[CII]} = 0.50$.

The left panel of Figure \ref{fig:CGMfraction} indicates that there is a strong positive correlation ($\rho = 0.9$) between $\rm f_{[CII]}$ and $r_{\rm p}$, meaning that merging systems with larger projected separations appear to have a larger $\rm f_{[CII]}$ compared to systems with smaller $r_{\rm p}$, where a clear separation between the emission coming from the diffuse component and that coming from the galaxies is more challenging. However, it could also be that galaxies that are interacting for a longer period of time result in a more carbon-rich envelope (see Appendix \ref{app:B}). To investigate this possibility we need to have information on the time evolution of the system, which is beyond the scope of the present work.

We also find an anti-correlation ($\rho = -0.8$) between the fraction of [CII] emission coming from the inner CGM and the mass ratio of the galaxies: the larger $\mu_\star$ the smaller $\rm f_{[CII]}$. This may suggest that the dynamical interaction between mergers where galaxies have similar masses ($\mu_\star \sim 1$) results in stronger tidal stripping and, as a consequence, in a more polluted CGM. Finally, we analyzed the relation between $\rm f_{[CII]}$ and $\mu _{[\rm CII]}$ (right panel), finding a mild correlation, $\rho = 0.5$, which suggests that the fraction of [CII] emission from the diffuse envelope does not depend on the ratio between the $\rm L_{[CII]}$ associated with the galaxies.

The system highlighted in gray in Figures \ref{fig:diffuseCGM} and \ref{fig:CGMfraction}, is DC\_873321. It is the only one for which we had to perform a two 2D Gaussian components fit instead of two \textit{single} 2D Gaussian models (see Section \ref{sec:OBSanalysis}), because of the difficulty in separating the emission from the galaxies and that of the diffuse envelope. 

To summarize, our observational analysis shows that a consistent part ($\sim 25 \%$, at least) of the [CII] emission of major merging galaxies identified in the ALPINE survey originates from a diffuse gas envelope around the merging galaxies or inner CGM. Also, we find tentative trends between the emission arising from the diffuse envelope and the physical properties of the merging systems, which provide some indication of the possible mechanisms that are responsible for the presence of a diffuse metal-enriched cold gas component surrounding these galaxies, such as small satellite companions, gravitational interaction resulting in tidal stripping and outflows activity.

In the next section, we compare the above findings and observational sample with synthetic merging systems at similar redshift selected from a cosmological hydrodynamical simulation, to gain a more in-depth understanding of the physical processes shaping the origin and metal enrichment of their inner CGM.

\section{Comparison with simulations}
\label{sec:simulations}
Intrigued by the results obtained from ALMA observations, which suggest that a fraction of the [CII] emission originates from the gas between merging galaxies and that different mechanisms (i.e. outflows, star-forming satellites, and tidal interactions) can be responsible for such emission, we employ cosmological simulations to gain more insights into the nature of [CII] emission. In what follows, we first provide a brief presentation of the simulations that will be used for our analysis. These have been thoroughly presented in \citet{DiCesare+2023} and have been run using the hydrodynamical code \texttt{dustyGadget}
\citep{Graziani+2020}. We then illustrate our analysis of the simulation results aimed at interpreting the nature of the observed [CII] envelopes. To this aim, we identify simulated galaxies with properties similar to the observed systems, and we use the carbon mass in their cold ISM medium (T $\lesssim 10^4 $ K) as a proxy for their [CII] emission. Our goal is to investigate whether the simulated galaxies follow the relations presented in Section \ref{sec:diffEmission}. We remark here that our qualitative analysis\footnote{Radiative transfer codes are indeed needed to estimate the proper [CII] luminosity from synthetic galaxies, see discussion in Section \ref{sec:discussion}.} is meant to provide \textit{indications} on the physical processes that may be responsible for the extended emission in merging systems, supporting the interpretation of ALMA observations.

\subsection{\texttt{dustyGadget}}
\label{subsec:dustyGadget}
In this section, we provide a synthetic summary of the key features of the cosmological hydrodynamical code \texttt{dustyGadget}, and we refer the interested reader to \citet{Graziani+2020} for a detailed description of the code, and to \citet{DiCesare+2023, Venditti+2023b, Venditti+2023} for its latest results. 

\texttt{dustyGadget} is an extension of the particle-based Smoothed Particle Hydrodynamics (SPH) code \texttt{Gadget-2/3} \citep{Springel2005,Springel+2021} which provides a self-consistent implementation of dust production and evolution, on top of the improvements to the chemical evolution module by \citet{Tornatore+2007a,Tornatore+2007b}, and to molecular chemistry and cooling by \citet{Maio+2007}. In particular, the gas chemical evolution scheme is inherited from \citet{Tornatore+2007a} (see also \citealt{Maio+2010}): it follows the metal release from stars of different masses, metallicity, and lifetimes. Yields depending on mass and metallicity are implemented for both PopII and PopI stars and stars with masses $\ge 40 \: \rm{M_{\odot}}$ are assumed to collapse into black holes and do not contribute to metal enrichment. PopIII stars with masses inside the range $140 \: \rm{M_{\odot}} \le \rm M_{\star} \le 260 \: \rm{ M_{\odot}}$ are expected to explode as pair-instability SNe (PISN), while those with masses outside the PISN range are assumed to collapse into black holes. We warn the reader that the formation of AGN is not modeled in our simulations, thus we do not account for mechanical or radiative feedback from accreting nuclear black holes. Moreover, as already stated, [CII] emission is not followed in the simulation, hence we use the carbon mass in the cold gas phase as a proxy for the emission (see discussion in Section \ref{sec:2systems}). Following \citet{Maio+2007}, the chemical network in \texttt{dustyGadget} also includes the evolution of both H and H$^+$, He, D, and primordial molecules by relying on the standard \texttt{Gadget} implementation of the cosmic UV background as a photo-heating mechanism, first introduced by \citet{Madau1996}. Star formation occurs in the cold gas phase, once the gas density exceeds a value of $n_{\rm th} = 132 \, h^{-2} \rm cm^{-3}$ (physical). The IMF of the stellar populations, each represented by stellar particles, is assigned according to their metallicity $\rm{Z_\star}$, given a gas critical metallicity $\rm{Z_{\rm crit}} = 10^{-4} \: \rm{Z_\odot}$, where $\rm{Z_\odot} = 0.02$ \citep{Grevesse&Anders1989}. When $\rm{Z_\star < Z_{\rm crit}}$ we adopt a Salpeter IMF \citep{Salpeter+1955} in the mass range $[100 - 500]\: \rm{M_\odot}$; otherwise, the stars are assumed to form according to a Salpeter IMF in the mass range $[0.1 - 100] \: \rm{M_\odot}$.

The production of dust grains by stars in SNe explosion and Asymptotic Giant Branch (AGB) stellar winds is described by means of a set of mass- and metallicity-dependent yields \citep{BianchiSchneider2007, Marassi+2014, Marassi+2015, Bocchio+2016, Ginolfi+2018, Graziani+2020, Pizzati+2020, Romano+2023}, which closely follow the chemical network adopted for metal yields \citealt{Tornatore+2007a}. Once the grains produced by stars are released in the ISM, they can undergo diverse physical processes that can alter their mass, relative abundances, chemical properties, charge, and temperature. Generally, it is assumed that the dust-to-light interactions (e.g. photo-heating, grain charging) change the thermodynamic and electrical properties of the grains (see for example \citealt{Glatzle+2019,Glatzle+2022}), but these processes have a negligible impact on the total dust mass ($\rm M_{dust}$) unless the grain temperatures reach the sublimation threshold ($\rm{T_{\rm d,s}} \gtrsim 10^{3}$~K). Other physical processes (i.e. sputtering and grain growth) can alter the total dust mass and the grain size distribution \citep{Draine+2011,Aoyama+2020}. In the version of \texttt{dustyGadget} that we adopt in this work, we only consider physical processes which alter the dust mass (i.e. grain growth, destruction by interstellar shocks, and sublimation)\footnote{We do not consider the evolution of grain sizes, and the grains are assumed to be spherical, with a static, average size of 0.1 $\mu$m \citep{Graziani+2020}.} in the hot and cold phases of the ISM \citep{SpringelHernquist2003}. The spreading of dust grains and atomic metals in the ISM, CGM, and IGM is done through galactic winds, which are modeled with a fixed initial velocity following the implementation of \citet{Tornatore+2010} - see \citet{Maio+2011} for the interplay between chemical and mechanical feedback in simulations. In \texttt{dustyGadget}, the initial velocity is fixed to 500 km/s, following typical outflow velocities observed in main sequence galaxies at $z\gtrsim4$ \citep{Gallerani+2018, Sugahara+2019, Ginolfi+2020b}. Finally, the identification of DM halos and their substructures is performed in post-processing with the AMIGA halo finder (AHF, \citealt{Knollmann+2009}). 

In this work we make use of \textit{one} of the eight statistically independent cosmological simulations (U6-U13) which have been analyzed in \citet{DiCesare+2023}, where we have investigated the redshift evolution of the SFR, stellar mass density, stellar mass function, and galaxy scaling relations, at $z \geq 4$, comparing our results with observational data. 

The physical setup of the runs and simulated cosmic volumes are the same as in \citet{DiCesare+2023}; in particular, we consider a cubic volume with $50\; h^{-1}$~cMpc ($\sim$ 74 cMpc) side length, with $2 \times 672^3$ particles, corresponding to a mass resolution of $5.2 \times 10^7 \, \rm M_\odot$ for DM particles, and of $8.2 \times 10^6 \, \rm M_\odot$ for gas particles. Using the current version of \texttt{dustyGadget} we can resolve down to the scale of a giant molecular cloud (a few tens up to hundreds pc), but for the purpose of this work, we are only interested in scales of the order of tens kpc.

Finally, the simulations adopt a flat $\Lambda$CDM cosmology with $\Omega _{\Lambda} = 0.6911$, $\Omega_{\rm m} = 0.3089$, $\Omega _{\rm b} = 0.0486$ and $h=0.6774$ \citep{PLANK+2016}. The simulations start at $z=100$ assuming neutral pristine gas and evolve all particle components down to $z = 4$. In the following, whenever we refer to units we mean \textit{physical} one, unless otherwise stated. 

\subsection{Identifying galaxy mergers}
\label{sec:findMergers}
\begin{figure}
\centering
\includegraphics[width=0.5\textwidth]{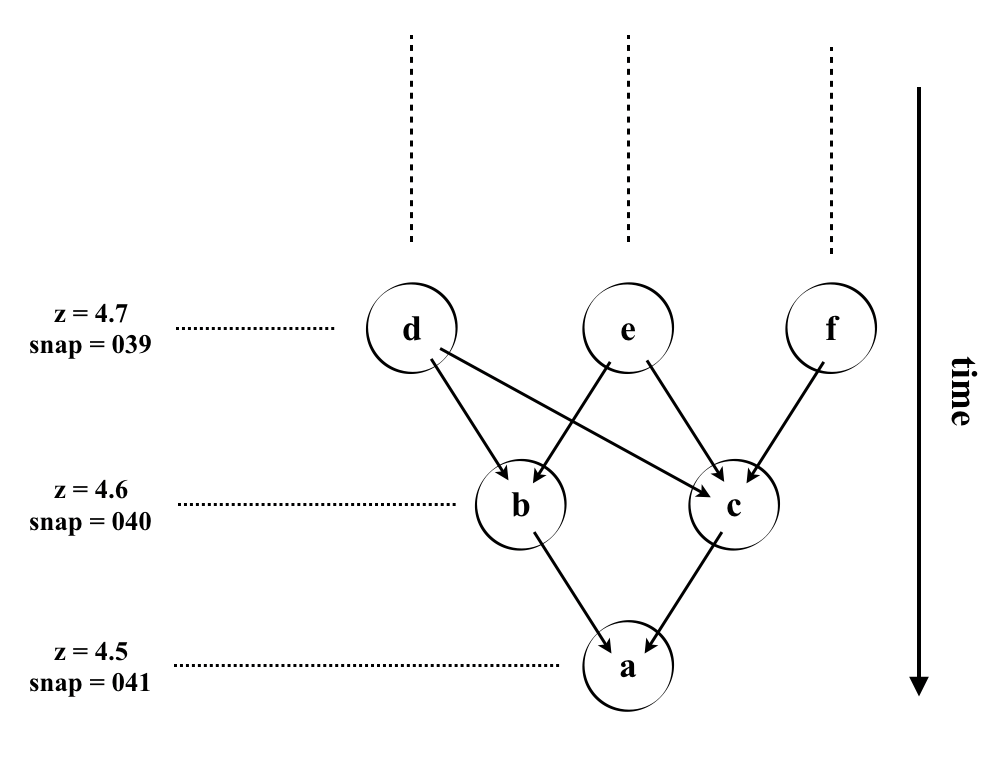}
\caption{Illustration of an example of merger tree reconstruction. Starting from galaxy "a" at z = 4.5 (snap = 041), its progenitors at the previous snapshot are "b" and "c". Merger trees have been reconstructed for each of all simulated galaxies at $z = 4.5$ with masses $M_\star \geq 10^{10} M_\odot$, up to $z = 5.1$ (snap = 035). This allows us to reconstruct the mass assembly history of the simulated galaxies, identifying those systems that undergo major mergers, defined as galaxy interactions where the relative stellar mass ratio of the merging pair is $1 < \mu_\star < 4$ (see text).}
\label{fig:MTree_sketch}
\end{figure}
\begin{table*}
    \centering
    \caption{Properties of the 8 simulated systems classified as major mergers and the 2 galaxies which are not in major mergers. For each redshift, $z$, we show the system ID, the distance obtained averaging over distances between merging galaxies on the x,y,z projections, $r_m$ [kpc], the carbon mass ratio, $\mu_C$, the relative mass ratio of the merging pair, $\mu_\star$, the total stellar mass, M$_\star$ [M$_\odot$] and SFR [M$_\odot$ yr$^{-1}$]. In the last two rows, we report integrated properties for non-interacting systems which we use together with merging systems, see discussion in Section \ref{sec:discussion}.}

    \begin{tabular}{c|c|c|c|c|c|c}        \hline
    system ID & $z$ & $r_m$ [kpc] & $\mu _C$ & $\mu _\star$ & Log($\rm M_\star/M_\odot$) & Log($\rm SFR/ M_\odot yr^{-1}$)  \\
    \hline
    \multicolumn{7}{c}{merging galaxies} \\
    \hline
    \hline
    H2 & 4.495 & 11.3 & 2.0 & 4.1 & 10.44 & 1.84 \\
    H4 & 4.495 & 16.3 & 2.3 & 1.5  & 10.55  & 1.96 \\
    H5 & 4.495 & 6.3 & 1.3 & 1.9  & 10.49 & 1.83  \\
    H6 & 4.495 & 5.3 & 1.3 & 1.9 & 10.43 & 1.81 \\
    H10 & 4.495 & 10.3 & 3.2  & 1.3 & 10.51 & 1.90 \\
    H25 & 4.495 & 40.3 & 2.8 & 2.3 & 10.30 & 1.70 \\
    H28 & 4.495 & 32.3 & 3.6 & 4.1 & 10.27 & 1.74 \\
    H0 & 4.988 & 13.3 & 4.6 & 1.5 & 10.58 & 2.17 \\
    \hline
    \multicolumn{7}{c}{non-interacting galaxies} \\
    \hline
    \hline
    H7 & 4.495 & - & - & - & 10.43 & 1.84 \\
    H29 & 4.495 & - & - & - & 10.58 & 1.85 \\
    \end{tabular}
    \label{tab:simGal}
\end{table*}
\citet{DiCesare+2023} already compared the integrated properties (SFR, $\rm M_\star$, $\rm M_{dust}$) of synthetic galaxies from \texttt{dustyGadget} with those observed in the ALPINE survey, finding a remarkable agreement between the two (see for example their Figure 3 and Figure 8).

In this work, we start from the galaxy catalogs generated by the simulation and select synthetic merging galaxies whose physical, integrated properties resemble the observed ones, as described in Section \ref{targets}. We analyze the simulation snapshots (hereafter referred as snap) from snap = 041 ($z = 4.495$) to snap = 035 ($z = 5.098$), which encompass the redshift range of the observed systems ($z = 4.5 - 5.1$, see Table \ref{tab:OBStargets}), and a physical time interval of $\Delta \rm t_{H} = (1.338 - 1.145)$ Gyr = 193 Myr, for our assumed cosmology. Hence, we end up having six galaxy catalogs, with $\Delta z$ = 0.1, containing simulated systems with masses in the range M$_\star \sim 10^8 - 10^{11} \, \rm M_\odot$ ($\sim$ 14000 objects at snap = 041, and $\sim 9400$ at snap = 035). Among these, we consider only galaxies with stellar mass M$_\star \geq 10^{10} \, \rm M_\odot$, to match the properties of the observed systems (see Table \ref{tab:OBStargets}), ending up with $\sim$ 40 galaxies at snap = 041 and $\sim$ 15 galaxies at snap = 035. Within these smaller samples of objects, which match the observed systems in terms of redshift and stellar mass, we must identify possible galaxy mergers.

Mergers are identified considering galaxies at $z \simeq 4.5$ and reconstructing their merger trees (i.e. their past assembly histories in gas, dark matter, and stellar particles) up to $z \simeq 5.1$. Figure \ref{fig:MTree_sketch} shows a schematic view of a merger tree reconstruction: we start at $z$ = 4.495 (snap = 041) with a galaxy dubbed "a", and go backward in time up to $z$ = 5.1 (snap = 035); at each intermediate redshift step, we identify the ancestors of "a" (i.e. halos "b" and "c" contributing particles to "a" at $z$ = 4.6). Once we have reconstructed the merger tree of each galaxy, we can then follow its evolution forward in time and characterize the origin of its mass assembly, whether this is the result of a major merger, smooth accretion, or minor mergers. 

In particular, we follow the same criterion adopted for the observed systems, and we assume that two ancestors undergo a major merger when their relative stellar mass ratio (defined as the ratio between the most massive and the least massive) is $1 < \mu_\star < 4$.

After applying this condition and checking that we are not double counting merging episodes\footnote{Double counting can happen if we count as different mergers those happening during the evolution of the same galaxy among the same ancestors across different redshift steps of its merger tree.}, we end up with 8 unique candidates (7 at $z \simeq 4.5$ and one at $z \simeq 5.0$), which undergo major mergers during their history. In Table \ref{tab:simGal}, we provide a summary of their physical properties, where each row corresponds to one of these synthetic major mergers. In particular, we report their IDs, redshift, the mean projected (r$_{ m}$)\footnote{This distance has been calculated averaging over the projected distances between the merging galaxies on the x,y,z planes.} distance in kpc, the relative carbon mass ratio of the merging pair ($\mu _C$), the relative stellar mass ratio of the merging pair, $\mu _\star$, the total M$_\star$ in M$_\odot$ and the SFR in M$_\odot$ yr$^{-1}$.

Compared to the observed sample (see Table \ref{tab:OBStargets}), the simulated systems appear to span a similar range of properties and can be considered as good synthetic analogs of the systems that we have analyzed in Section \ref{sec:observations}. This sample of synthetic galaxies can help us interpret the observed trends that we presented in Section \ref{sec:diffEmission}. As an illustrative example, in the next section, we discuss the time evolution of the simulated system H4 at $z = 4.495$, which provides the best synthetic analog of vc\_5100822662.

\begin{figure*}[!ht]
\centering
\includegraphics[scale=0.28]{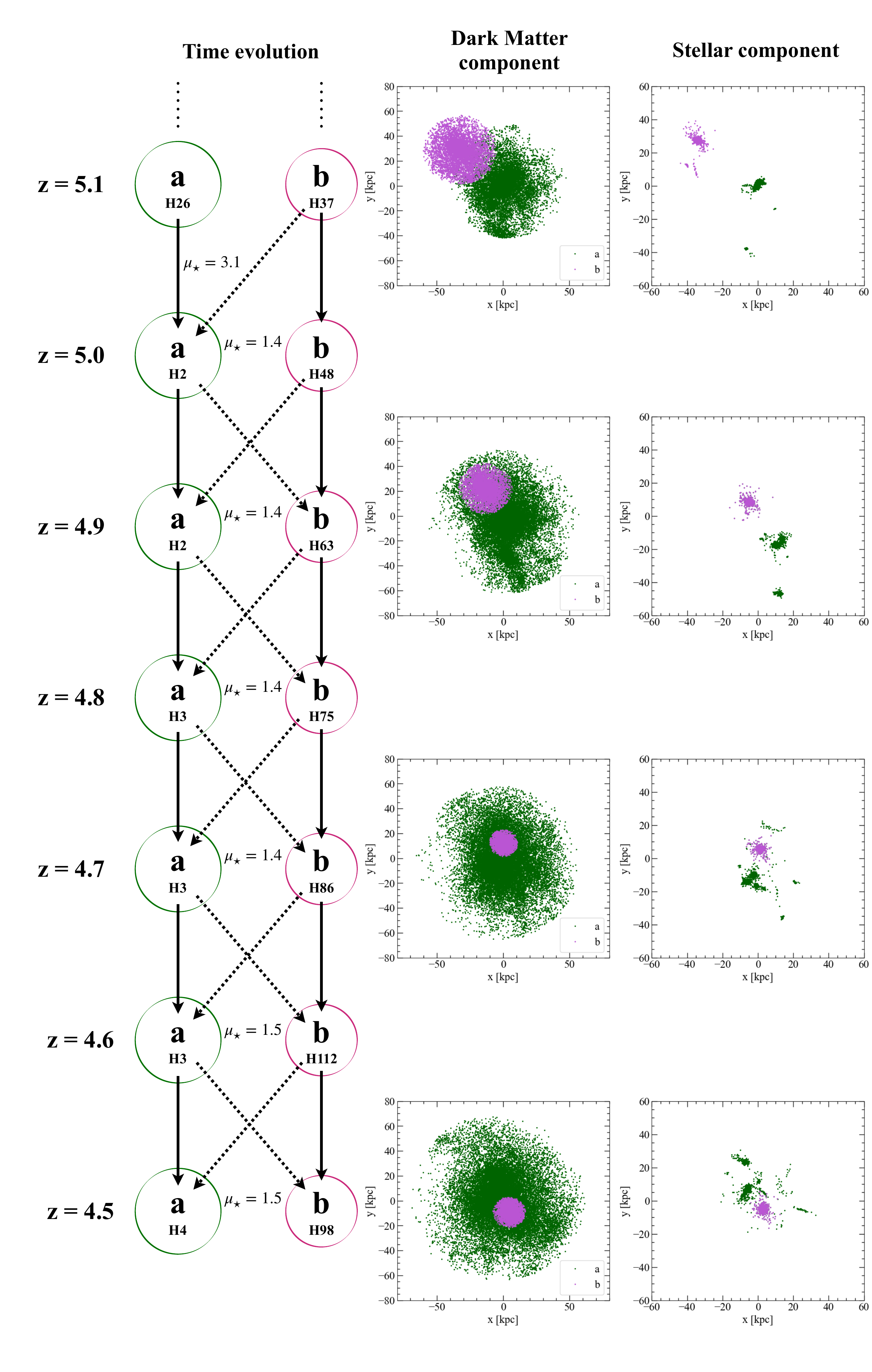}
\caption{\textit{Left:} History of two merging galaxies, H4 and H98 (dubbed as "a" and "b"), reconstructed along their past merger tree, starting from $z = 4.5$ up to $z = 5.1$. At each redshift, we show both the main progenitor (black solid line) and secondary progenitor (black dashed lines), we also report the stellar mass ratio $\mu _\star$ of the primary (green) and secondary (magenta) merging galaxies. In the \textit{Middle and Right} panels we show the projection on the (x,y) plane of the merging galaxies as seen in dark matter and baryonic matter components, respectively. For the sake of clarity, we only show the maps at four redshift steps ($z$ = 5.1, 4.9, 4.7, and 4.5, from top to bottom) and galaxies are color-coded as in the merger tree.}
\label{fig:MTree_example}
\end{figure*}

\subsection{Time evolution of a representative merger}
\label{sec:mergertree}
The simulated system H4 at $z = 4.495$ and the observed system vc\_5100822662 (hereafter, vc\_662) at $z = 4.5210$ have stellar mass and SFR which are consistent within the errors: Log($ \rm M_\star/M_\odot$) = 10.6 (H4) and Log($ \rm M_\star/M_\odot$) = 10.4$^{+0.1}_{-0.1}$ (vc\_662); Log($\rm SFR / M_\odot \; yr^{-1}$) = 2.0 and Log($\rm SFR / M_\odot \; yr^{-1}$) = 2.0$^{+0.2}_{-0.2}$, respectively. Moreover, H4 and vc\_662 have similar stellar mass ratios $\mu _{\rm H4}$=1.5 and $\mu _{\rm vc\_662}$=1.7, and projected distance of 16 kpc and 11 kpc respectively. It is important to keep in mind that from the simulation, we not only have information on projected distances but also on the physical (3D) distance between the merging galaxies, which we compute to be $r$ = 21 kpc for this synthetic system.   

Figure \ref{fig:MTree_example} shows the merger tree reconstruction for H4 (see Table \ref{tab:simGal} for its integrated physical properties) from $z = 4.5$ to $z = 5.1$, together with maps (80 $\times$ 80) kpc and (60 $\times$ 60) kpc showing respectively the (x,y) projection of the dark matter particles and stellar particles. These maps are centered respectively in the center of mass of the dark matter halo and of the baryonic component. Here we dubbed the merging galaxies H4 and H98 at $z = 4.5$ as "a" and "b", respectively, so that it is easier to follow their past history along the merger tree reconstruction. In the illustration provided, we connect with solid black lines each component of the merging pair to its main galaxy progenitor at the preceding redshift\footnote{Main progenitor is the galaxy which contributes the most, in terms of stellar mass, to the descendant at the subsequent redshift.}: as an example, the galaxy dubbed as "a" at $z = 4.6$ (H3) is considered as the main progenitor of galaxy "a" at $z = 4.5$ (H4). At each redshift, we also indicate the $\mu _\star$ of the galaxies undergoing the major merger. For the sake of clarity, the projections of the dark matter and baryonic components in the middle and right panels of Figure \ref{fig:MTree_example} are shown only at four redshifts, and particles belonging to each merging component are color-coded to match those in the corresponding merger tree. In the following section, we discuss how the spatial distribution of the gaseous component is generated. 

Looking at the dark matter and baryonic matter visualizations in the middle and right panels of Figure \ref{fig:MTree_example}, it is interesting to note how dark matter halos merge and coalesce in a shorter time compared to galaxies themselves (i.e. to the baryonic components). Indeed, by $z = 4.5$, the dark matter halos of the two systems can be considered as a single virialized object, while the baryonic components are still largely spatially segregated, and undergoing the merger. Moreover, as soon as the gravitational interaction between galaxies becomes stronger (i.e. they move closer), galaxies appear to be more clumpy and rich in satellites and undergo disruption events which lead to filamentary structures by $z = 4.5$. In addition to this, we also have to consider that the merger between the dark matter halos of the galaxies may lead to the formation of overdense regions which result in dense clouds where star formation can be triggered. 

All this qualitative information coming from the simulation provides us with fundamental indications on the evolution of merging pairs, inaccessible using only observations, and helps us interpret the results from the observational part of this work. In fact, simulated galaxies have higher spatial resolution compared to ALPINE observations, and we are able to resolve small satellite galaxies and filamentary structures. This allows us to infer the origin of the enriched diffuse gas surrounding the galaxies, which powers the observed [CII] emission coming from the inner CGM of merging galaxies: whether this is due to unresolved satellites galaxies, filamentary structures, outflows or maybe a combination of the three. In the following, we will explore these options in a more quantitative way.

\subsection{Dependence of galaxy properties and their CGM on the mass ratio of merging pairs}
\label{sec:2systems}

\begin{figure*}
\centering
\includegraphics[scale=0.42]{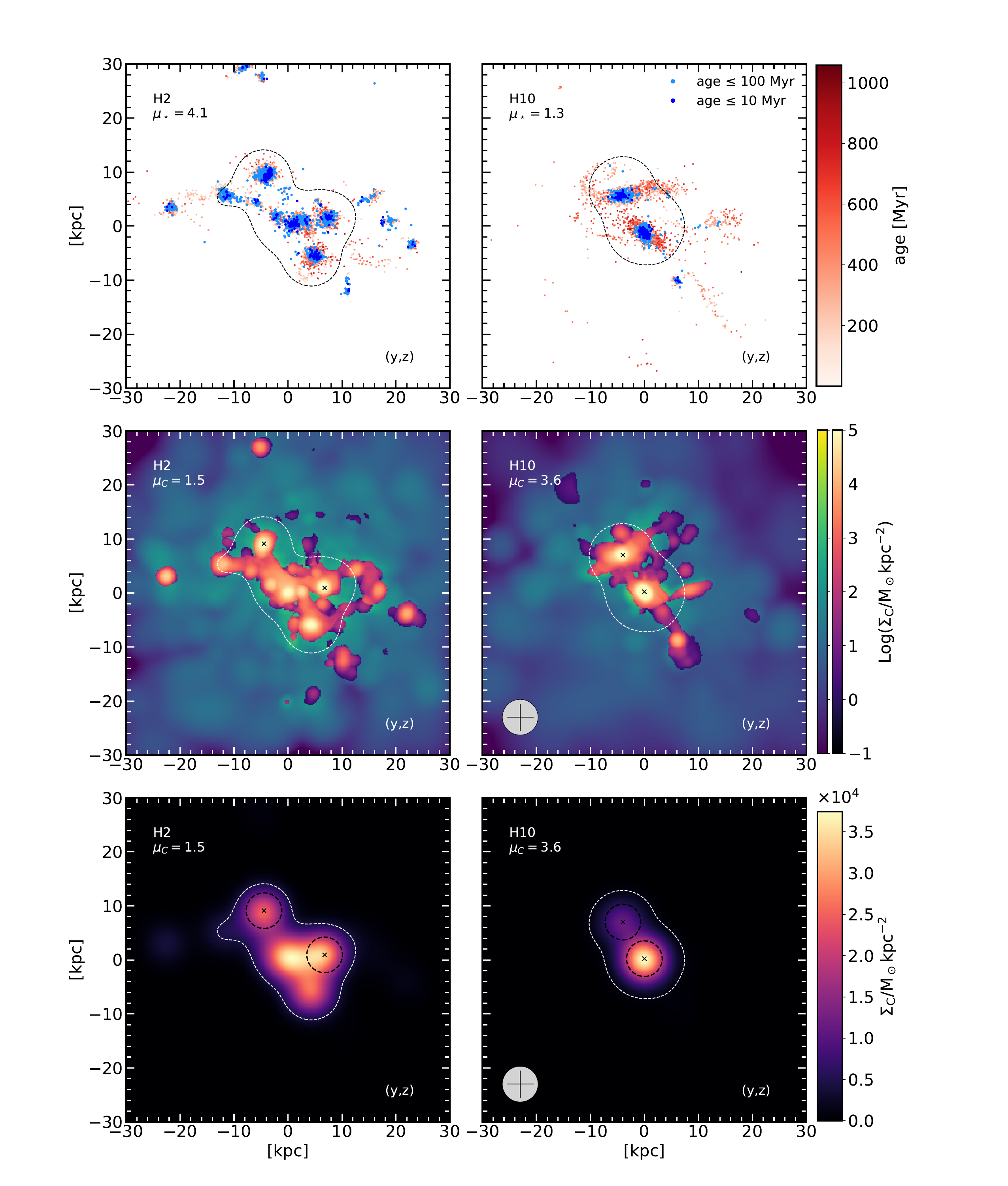}
 \caption{Properties of two examples of synthetic merging systems with high ($\mu _\star$ = 4.1, H2, left panels) and low ($\mu _\star$ = 1.3, H10, right panels) stellar mass ratios at $z = 4.5$. \textit{First row:} maps of the stellar surface density projected along a line-of-sight parallel to the $x$-axis. Stellar particles are color-coded according to their age, with age $\leq 10$ Myr in dark blue, age $\leq$ 100 Myr in light blue, and 100 Myr < age $ \leq 1$ Gyr in red color scale. \textit{Second row:} same as before but for the carbon surface density distribution of H10 ($\mu _C = 3.6$) and H2 ($\mu _C = 1.5$), respectively. The green color map shows the \textit{total} carbon surface density distribution and overplotted on that, in orange tones, is the carbon distribution in the \textit{cold gas phase} (i.e. with T$ < 5 \times 10^4$ K). \textit{Third row:} carbon in the cold gas phase surface density distribution once we convolve the original simulated map with the ALMA PSF. Black ellipses indicate the FWHM-based apertures used to extract the carbon content associated with each merging galaxy. In the bottom left corner is a beam of $\sim 1"$ as a reference. Black (first row) and white (second and third rows) dashed contours indicate the positive significant level at 2$\sigma$.}
\label{fig:stars+C}
\end{figure*}

In this section, we focus on two synthetic candidates (H2 and H10) at $z = 4.5$ which have been selected as their merging pairs are characterized respectively by the highest ($\mu _\star = 4.1$) and lowest ($\mu _\star = 1.3$) stellar mass ratios in our sample. Our aim is to investigate whether systems with extreme $\mu _\star$ may be characterized by different morphological properties of the stellar and gas components, on scales comparable to the diffuse [CII] emission component quantified in Section \ref{sec:diffEmission}.

Figure \ref{fig:stars+C} shows the stellar particle distributions (first row) and carbon surface density maps ($\Sigma _{\rm C}$, second and third rows) for H2 and H10 projected on the plane perpendicular to the line-of-sight along the $x$-axis. Specifically, in orange/violet tones is the carbon in regions with gas temperature T $ < 5 \times 10^4 $ K, used as a proxy for L$_{\rm [CII]}$ - see following discussion.  

In general, to characterize the gas and metals spatial distributions we project gas and metal particles onto Cartesian grids with 256 or 512 cells/side, and each particle contribution is weighted with the SPH kernel adopted in \texttt{dustyGadget}. These maps are centered either in the halo mass center (H2, H4, H5, H6, H10, H0) or in the center of mass of the merging system (H25, H28), depending on how diverse these two values are, while the side of the Cartesian grid is chosen to capture the whole merging system. Specifically, we chose a box side of 60 kpc (256 cells/side) for all the synthetic candidates but for H25 and H28 for which we adopted a side length of 100 kpc (512 cells/side). Both grids have a spatial resolution (pixel dimension) of $\sim$0.2 kpc, thus a factor of $\sim$ 20-30  more resolved than the ALPINE-ALMA beam.

As for the panels showing stellar particles (scatter plots), color-coded in red are their ages, while, we highlight in dark blue stellar populations with ages $\leq$ 10 Myr and in light blue those with ages $\leq$ 100 Myr. This is done in order to characterize younger populations and star-forming regions (blue) in contrast to more mature stellar populations in red. Observations interpreted with photo-ionization models suggest that the bulk of [CII] emission is coming from neutral atomic gas clouds in photo-dissociation regions (PDRs) surrounding young stars \citep{Hollenbach+1999}, thus we expect that a consistent part of the carbon mass predicted by \texttt{dustyGadget} simulations will be associated with young stellar populations (ages $\leq$ 10 Myr) - see discussion in Section \ref{sec:discussion}.  

In the second row of Figure \ref{fig:stars+C}, we color in blue/green tones the \textit{total} carbon surface density, while in orange/violet the carbon surface density found in regions with gas temperature T $ < 5 \times 10^4 $ K \footnote{This temperature is the hydrodynamical one, thus it does not take into account cooling effects which may arise once we apply radiative transfer codes.}. This cut has been done as the [CII] emission is expected to be associated with cold-warm gas phases, and at higher temperatures we expect the C atoms to be in higher ionization states. 

The third row of Figure \ref{fig:stars+C} shows the surface density distribution of the carbon associated with gas at a temperature below the adopted cut (hereafter "cold" gas phase) once it is convolved with the ALMA PSF (i.e. we applied a Gaussian smoothing with FWHM $\sim 1"$). This step is needed since we want to compare the results from simulations with those from observations (see Section \ref{sec:statProperties}). The black crosses mark the center of the two interacting galaxies and the gray circle in the bottom left corner shows a reference beam of $\sim 1 "$.      

The comparison among two merging systems in Figure \ref{fig:stars+C} highlights the different environments these systems belong to. In fact, H2 is characterized by a dominant component and several smaller satellites ($\mu _\star \sim 4$), while H10 is characterized by an almost equal mass merging pair ($\mu _\star \sim 1$). This translates into different stellar and metal-enriched gas spatial distributions in the two cases.

Also, it is interesting to note how diverse scenarios appear when comparing carbon maps with and without the convolution with ALMA PSF, with the latter lacking the detailed spatial distribution inferred from the simulation.    

\begin{figure*}
\centering
\includegraphics[scale=0.3]{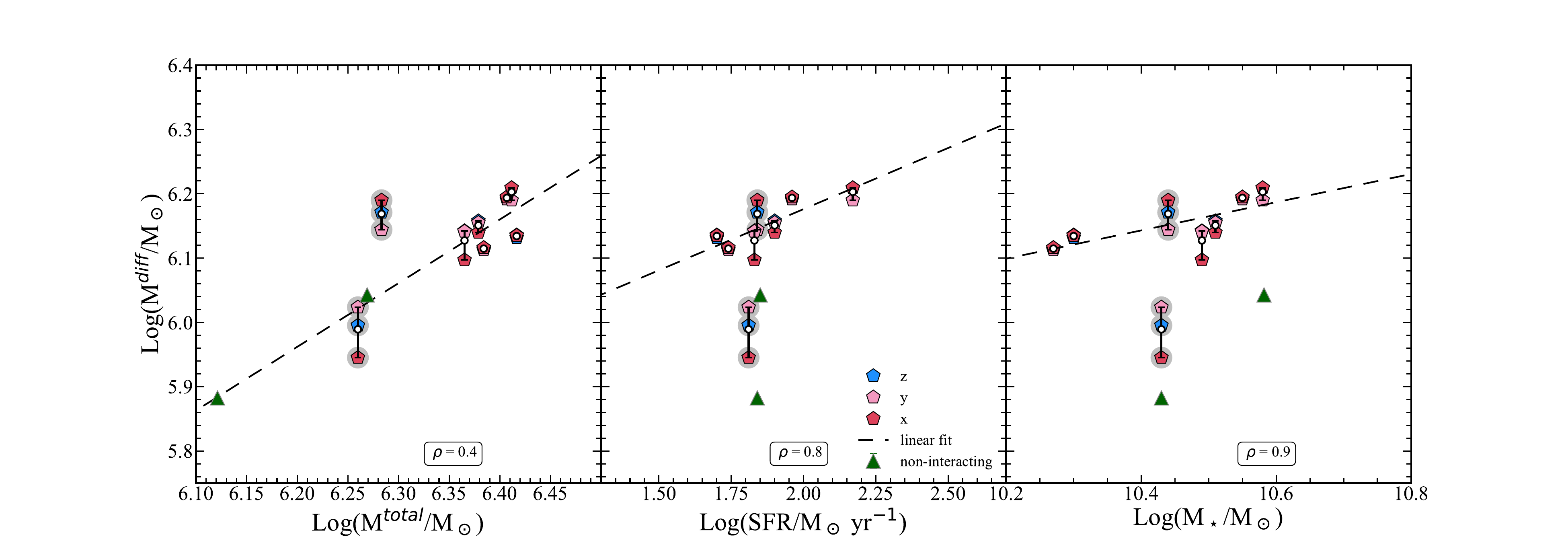}
 \caption{Trends between the carbon mass in the cold gas phase ($\rm T < 5 \times 10^4 K$) coming from the diffuse halo around merging systems, M$^{\rm diff}$, and their integrated physical properties. Different colors refer to projected quantities along different lines of sight parallel to the $z$ (blue), $y$ (pink), and $x$ (red) axis. In each panel, white dots are $\rm M^{avg}$ (i.e. the average among the three projections), the black dashed line is the linear fit once we consider $\rm M^{avg}$, and $\rho$ is the Spearman coefficient corresponding to the analyzed relation. \textit{Left:} M$^{\rm diff}$ is shown as a function of the total carbon mass in the cold gas phase (M$^{\rm total}$). \textit{Middle and Right panels:} M$^{\rm diff}$ as a function of the total SFR and total M$_\star$ of the merging systems. Highlighted in gray are H2 and H6, which have peculiar structures, and for this reason, they have not been included in the fit. Green triangles are single non-interacting galaxies, see the end of this section for discussion.}
\label{fig:diffuse_sim}
\end{figure*}

\begin{figure*}
\centering
\includegraphics[scale=0.3]{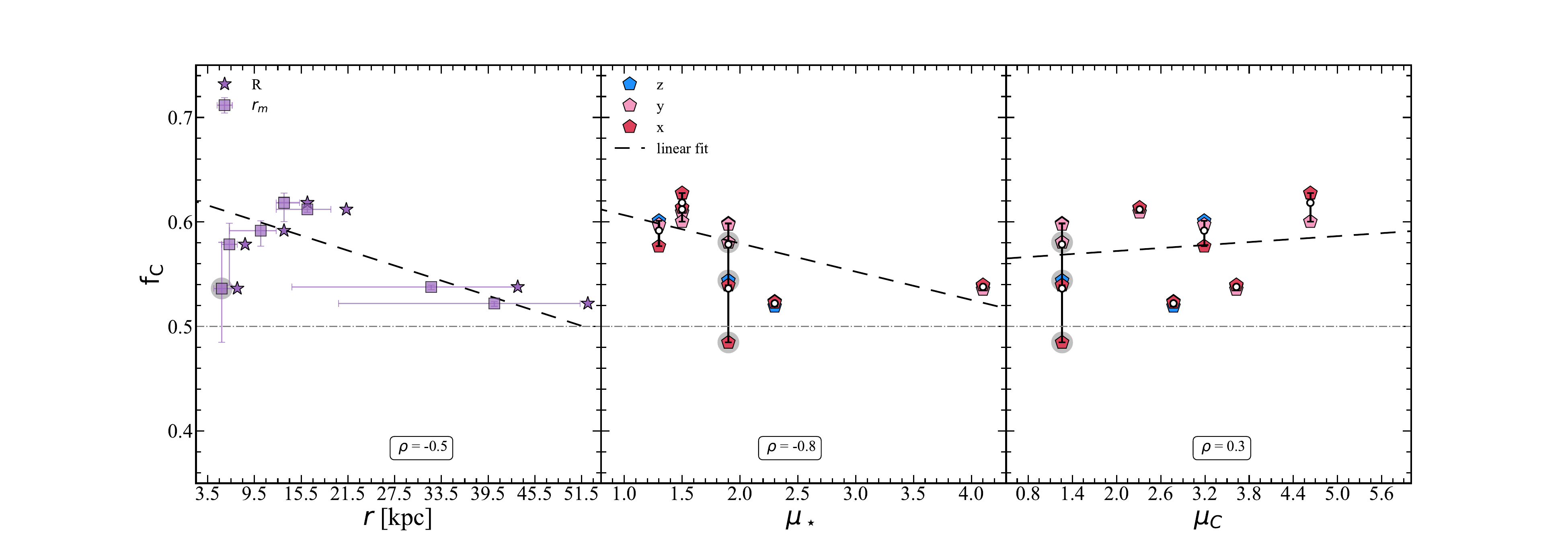}
 \caption{Trends between the fraction of carbon mass in the cold gas phase ($\rm T < 5 \times 10^4 K$) in the diffuse halo (f$_C$) and the distance ($r$) between the galaxies (\textit{Left}), the stellar mass ratio ($\mu_\star$, \textit{Middle}), and the carbon mass ratio ($\mu_C$, \textit{Right}). In particular, in the \textit{Left} panel we show $\rm f^{avg}$ (i.e. the average among the three projections) as a function of both the distance averaged on the three projections (r$_m$, violet squares) and the physical distance between the galaxies (R, violet stars). In the \textit{Middle and Right} panels different colors refer to f$_C$ estimated by projecting the carbon surface density along lines-of-sight parallel to the $z$ (blue), $y$ (pink), and $x$ (red) axis. In each panel, white dots are $\rm f^{avg}$, the black line is the linear fit considering $\rm f^{avg}$ versus r$_m$, $\mu_\star$, $\mu_C$ respectively. We also specify the Spearman coefficient ($\rho$) of each relation. H2 is not included in these plots as this system is characterized by an out-of-scale value of f$_C$ = 0.8 and H6 is highlighted in gray. The horizontal gray dot-dashed line shows the value of f$_C$ = 0.5.}
\label{fig:fraction_sim}
\end{figure*}
\subsection{Statistical properties}
\label{sec:statProperties}
Here we analyze the relations between the carbon mass - in the cold phase - present in the diffuse envelope around merging galaxies and the integrated properties of such systems, to identify possible trends. The aim is to derive scaling relations that can be compared to the observational ones (see Figure \ref{fig:diffuseCGM} and Figure \ref{fig:CGMfraction}) using the carbon in the cold gas phase as a proxy for [CII] luminosity. In fact, once similar relations are recovered, predictions from \texttt{dustyGadget} can help us interpret the observations and help us gain insights into the nature of the extended emission.

From now on, whenever we talk about the carbon mass we refer to the M$_C$ in the cold phase, T $ < 5 \times 10^4 $ K, orange/violet tones in Figure \ref{fig:stars+C}. As previously stated, this cut in temperature is necessary in order to highlight the regions from which we expect [CII] emission to be more likely. Briefly, we generate the gas and carbon 3D distributions, apply the cut in temperature to these cubes, and then project this result on planes perpendicular to the lines of sight parallel to the $z$, $y$, and $x$-axis. We repeat this procedure to all our selected synthetic galaxies (see Table \ref{tab:simGal}) and, once we have an image for each candidate and its three plane projections, we convolve the image with the ALMA PSF and fit a 2D Gaussian model to it, retrieving morphological information such as the coordinate of the main M$_C$ clumps that are associated with the major merging galaxies. Then, we use the mean value of all the FWHM-based apertures adopted for the observational part as the standard aperture to be used in simulations (black dashed circles in Figure \ref{fig:stars+C}, last row) and associate all the carbon inside the aperture as belonging to the galaxy.        

Knowing the coordinates of the center and the dimensions of all the apertures, for each 2D projection (lines of sight, $i = z,y,x$), we can distinguish between the carbon mass associated with the galaxies - sum of the carbon mass inside the apertures - and the total one - sum over the entire plane projection. At this point, we can define the M$_C$ associated with the diffuse halo as :

\begin{equation}
\rm    M^{\rm diff, i} \equiv M^{\rm total} - M^i
\end{equation}

where we dropped the subscript C. The fraction of M$_{C}$ in the envelope around merging systems (hereafter, $\rm f_{C}$) is:

\begin{equation}
   \rm f_{C}^i \equiv \frac{M^{\rm diff, i}}{ M^{\rm total}}.
\end{equation}

Figure \ref{fig:diffuse_sim} shows the relation between M$^{\rm diff}$ and some integrated properties of the systems, such as the total carbon mass, SFR, and stellar mass. For each relation, we calculate the Spearman coefficient to understand how reliable the suggested trend is. Moreover, different colors correspond to projected quantities along lines of sight parallel to the $z$ (blue), $y$ (pink), and $x$ (red) axis, and projections associated with the same system are connected by gray vertical lines quantifying the variation of M$^{\rm diff}$. 

The black dashed line is the result of the linear fit once we consider the carbon mass value in the diffuse halo averaged on the three projections (hereafter, $\rm M^{avg}$ shown as white dots in the figures) and the corresponding x-axis quantity. Finally, highlighted in gray are H2 which has a peculiar morphological structure rich in clumps and satellites (see left panels of Figure \ref{fig:stars+C}), and H6 whose galaxies are closely interacting (see Appendix \ref{app:A}). For these reasons, they are considered as not belonging to the fiducial simulated sample and have not been included in the fit.

The figure shows a positive $\rho =$ 0.4 and strong positive $\rho =$ 0.8, 0.9 correlations respectively in the first, second, and third panels. These results suggest that their inner CGM is characterized by a significant carbon mass, which appears to be well correlated with the SFR and stellar mass of the systems, indicating the important role that current and past star formation activity has in enriching the inner CGM with outflows from the merging pairs and smaller satellite systems. This conclusion strengthens the tentative trends between diffuse [CII] emission and star formation rate/stellar mass of merging systems ($\rho$ = 0.3 and 0.5, respectively) we found in Section \ref{sec:diffEmission}.

In Figure \ref{fig:fraction_sim} we explore the dependence between the fraction of carbon mass in the diffuse halo ($\rm f_C$) and r$_m$, $\mu _\star$ and $\mu _C$. In the middle and right panels, different colors correspond to different projections, with those associated with the same system connected by gray vertical lines. In each panel we specify the Spearman coefficient of the corresponding relation and the black dashed line is the linear fit once we consider the fraction of carbon in the diffuse envelope averaged on the three projections ($ \rm f^{avg}$) and the corresponding x-axis quantity. $\rm f^{avg}$ is shown with a violet square/star in the left panel and with white dots in the middle and right panels. The synthetic galaxy H2 has not been included in these plots being considered an outlier with $\rm f_C \sim 0.8$, then out of scale; H6, instead, has been highlighted in gray and not considered in the fits and computation of the Spearman coefficients.

On average, we find that, when considering the entire simulated sample, $\sim 59$\% of carbon mass resides in the envelope around merging systems. By excluding H2 which has been considered as an outlier being particularly rich in star-forming satellites (see left panels of Figure \ref{fig:stars+C}) this value becomes $\sim57$\%. Both cases are indicative of the presence of carbon in the cold gas phase in the inner CGM of the merging systems, according to \texttt{dustyGadget}.

In the left panel of Figure \ref{fig:fraction_sim} we show $\rm f^{avg}$ as a function of the mean projected distance (violet square) and physical distance (violet stars) between merging galaxies, with the physical distance being larger than the projected one in all the systems. The Spearman coefficient for the $\rm f^{avg}$ - r$_m$ relation is $\rho = -0.5$ suggesting a correlation with negative trend between these two quantities. This result, which points out a negative trend among $\rm f^{avg}$ - r$_m$, seems to disagree with what has been found in observations - see Figure \ref{fig:CGMfraction}. However, in simulations we are considering systems with mean distances r$_m$ up to $\sim 50$ kpc while, in the ALPINE sample, the projected distance goes up to $\sim 16$ kpc. This is further discussed in Appendix \ref{app:B}, where we show that a positive correlation among $\rm f^{avg}$ - r$_m$, similar to that found in observations, is recovered when we consider galaxies with $r_m \lesssim 16$ kpc, highlighting the role of interactions in polluting the diffuse medium.
 
The middle panel suggests an anti-correlation between the fraction of carbon mass in the diffuse halo and the stellar mass ratio of the merging galaxies, indicating that systems with $\mu _\star$ closer to unity have stronger interactions which result in more metal-enriched inner CGM. This may be a clue that dynamical interaction at high-$z$ can be an efficient mechanism for extracting material out of galaxies and mixing it in the CGM. Finally, the right panel shows $\rm f_C$ as a function of $\mu _C$, suggesting that there is not a correlation between the amount of $\rm M_C$ found in the diffuse halo and the ratio between carbon masses of each merging component. 

\begin{figure}
\centering
\includegraphics[width=0.5\textwidth]{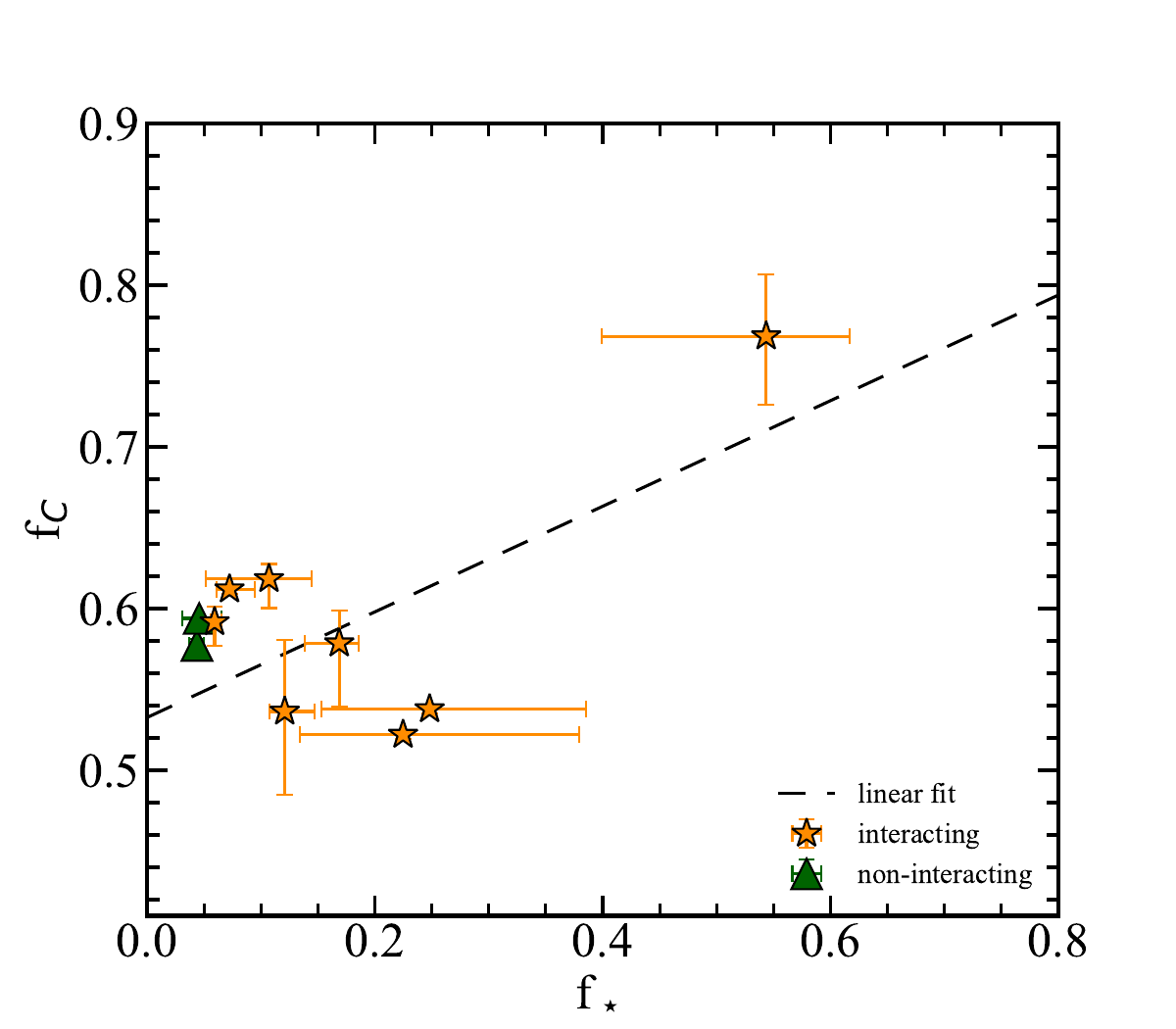}
\caption{Relation between the fraction of carbon mass (f$_C$) and the fraction of young - age $\leq$ 10 Myr - stellar populations (f$_\star$) present in the diffuse halo of interacting (orange stars) and non-interacting (green triangles) systems. We show the mean value of f$_C$ and f$_\star$ of all the simulated galaxies, together with the error bars representative of the variations in different projections and the linear fit.}
\label{fig:starF}
\end{figure}

\section{Discussion}
\label{sec:discussion}

In this section, we interpret the observations using our simulated galaxies and discuss the results obtained from the analyzed merging systems at $z \sim$ 4-6. First and foremost, it is important to bear in mind that the comparison between the results from observations and those from simulations is not straightforward. In fact, in simulations we consider the \textit{carbon mass} in the cold gas phase which is more likely to emit in [CII], but not the actual [CII] luminosity. In fact, to recover luminosity information from simulations we would need to couple our result with radiative transfer codes, such as \texttt{Cloudy} (for line transfer, \citealt{Ferland2017}) and \texttt{SKIRT} (for dust continuum, \citealt{Baes+2015}). These features will be implemented as a follow-up work. Thus, preliminarily, we assume that the carbon mass in the cold gas phase (T < $5 \times 10^4$ K) can be considered as a good proxy for [CII] luminosity, to guide the interpretation of the observational results illustrated in Section \ref{sec:diffEmission}.

On one hand, [CII] envelopes have been observed at first in non-interacting galaxies (e.g. \citealt{Fujimoto+2019, Ginolfi+2020b,Fujimoto+2020,Herrera-Camus+2021}), where they extend up to $\sim$ 10-15 kpc outside the galaxies. A possible explanation has been provided by \citet{Pizzati+2020, Pizzati+2023}, who adopt a semi-analytical model to interpret the presence of [CII] halos in high-$z$ galaxies concluding that they can be produced by ongoing (or past) starburst-driven outflows which transport carbon and other heavy elements in the CGM. Recently, \citet{Romano+2023,Romano+2024} analyzed \textit{Herschel} \citep{Pilbratt+2010} data of local dwarf galaxies finding that the observed [CII] emission - which size agrees with that measured for star-forming galaxies at $z > 4$ - can be attributed to the presence of galactic outflows. On the other hand, \citet{Ginolfi+2020a} found an even more extended (> 20 kpc) emission in one interacting system from the ALPINE sample, arguing that most of the detected circumgalactic emission is a consequence of the effect of gas stripping induced by strong gravitational interaction. 

In this work, by analyzing major merging systems in the ALPINE survey, we confirm the presence of extended [CII] envelopes around interacting galaxies, finding that around 25\% of the total [CII] emission comes from the medium between the galaxies. All the analyzed systems have an extended emission (mean value of $\sim$ 27 kpc) suggesting that the gravitational interaction between galaxies, which results in tidal tails, is responsible for their extended carbon-enriched envelope. This scenario has been investigated using the hydrodynamical simulation \texttt{dustyGadget} and comparing merging and not merging systems (see discussion in the following). Specifically, the anti-correlation between f$_C$ and $\mu _\star$ shown in Figure \ref{fig:fraction_sim} indicate that systems that interact more strongly result in a more enriched halo (i.e. higher f$_C$) - supporting the conclusion from \citet{Ginolfi+2020a}. Starburst-driven outflows, which spread metals outside the galaxies, are still at work in interacting systems. In fact, as shown in the middle panel of Figure \ref{fig:diffuse_sim}, there is a correlation between the carbon in the diffuse medium and the SFR of galaxies which can be associated with outflows. To quantify the importance of that mechanism compared to the others we need to look at galactic winds of interacting galaxies, which is something outside the aims of this paper.

Predictions from \texttt{dustyGadget} not only helped us corroborate and interpret the observational results, but we also used them to disentangle the different mechanisms that contribute to metal enrichment of the inner CGM of interacting galaxies. To do so, we consider SFR, $\rm M_\star$, and $\mu _\star$ as proxies for outflows, SF clumps, and tidal interactions, respectively, and study their contribution to the carbon-enriched envelope. 

As already mentioned, H2 is considered an outlier in the synthetic sample, because of its high f$_C$ which is due to its peculiar morphology rich in satellites/star-forming clumps. To further investigate this hypothesis (i.e. that star-forming clumps play a role in the observed [CII] diffuse emission), we study the relation among f$_C$ and the fraction of young, age $\leq$ 10 Myr, stellar populations (f$_\star$) that reside in the inner CGM of the systems. This quantity is defined as :
\begin{equation}
   \rm f_\star^i \equiv \frac{N_\star^{\rm total} - N_\star^i}{ N_\star^{\rm total}}.
\end{equation}
with $\rm N_\star^{\rm total}$ being the total number of young stellar populations and $\rm N_\star^i$ the sum of stellar particles associated with merging galaxies for each projection ($i = z,y,x$).

In Figure \ref{fig:starF} we show mean values for f$_\star$ and f$_C$ for interacting, in orange, and non-interacting, in green, systems; error bars on both axes give us an idea of the variations of f$_\star$ and f$_C$ among different projections. This figure shows a linear relation and positive trend between the fraction of carbon and that of young stellar populations in the diffuse halo, suggesting that clumps of star-forming regions, unresolved by ALMA, can play a role in enriching the gas envelope with carbon. Moreover, as shown in Figure \ref{fig:MTree_example}, the merger between dark matter halos happens on smaller timescales compared to those needed for the merger in the baryonic component and may lead to the formation of overdense regions at the periphery of merging systems triggering star formation and enriching their surrounding medium with carbon and other heavy elements. 

In addition to the previous analysis, by comparing results from major merging systems (dubbed as "interacting") with those from systems which are not in major mergers (dubbed as "non-interacting/single"\footnote{For simplicity we dubbed systems which are not undergoing major merger as "non-interacting/single", but it is important to keep in mind that these systems are interacting with their surrounding environments and other satellite galaxies, i.e. they are not isolated.}) we can have an estimate of the importance of gravitational interactions, which result in tidal stripping, for the enrichment of the inner CGM. To do so, we select two non-interacting galaxies with stellar masses and SFRs similar to the interacting synthetic candidates - see Table \ref{tab:simGal}. These two galaxies are also representative of other synthetic single galaxies at the same redshift that we find in our simulation. 

\begin{figure*}[!ht]
\centering
\includegraphics[scale = 0.5]{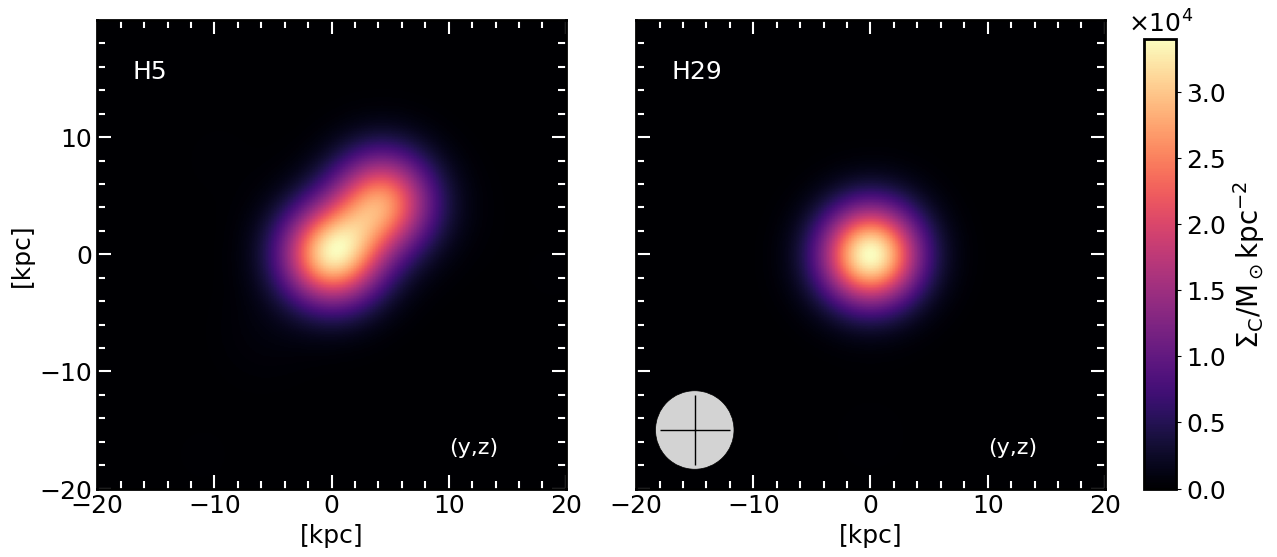}
\caption{As an example we show the surface density distribution of carbon in the cold gas phase once we convolve the original simulated map with the ALMA PSF, for interacting (left) and non-interacting (right) systems. The line-of-sight is chosen parallel to the $x$-axis, in the bottom left corner is a beam of $\sim 1"$ as a reference.}
\label{fig:singleInter}
\end{figure*}

In the single-galaxy case, we recover an extended carbon-rich envelope going up to $\sim $ 10 kpc, in agreement with what has been estimated by \citet{Fujimoto+2019,Fujimoto+2020,Ginolfi+2020b,Herrera-Camus+2021}. This is shown in Figure \ref{fig:singleInter}, and confirms that the diffuse halo in non-interacting galaxies is less extended than the one estimated for interacting systems (> 20 kpc).

In Figure \ref{fig:diffuse_sim} we compare M$^{\rm diff}$ in interacting (colored pentagons) and non-interacting (green triangles) galaxies. In general, we note that single galaxies have lower M$^{\rm diff}$ values than interacting systems, and also that they result in similar diffuse carbon masses no matter the adopted line-of-sight, indicating a more compact and regular morphology i.e. not disturbed by dynamical interactions. The first panel of Figure \ref{fig:diffuse_sim} suggests that both interacting and non-interacting systems follow a similar trend when comparing M$^{\rm diff}$ and the total carbon mass. On the one hand, the relation between the diffuse carbon mass and the SFR (second panel) shows that for non-interacting systems, the ongoing SFR is not a good proxy for the amount of carbon mass in the inner CGM. We also note that H6, which is an interacting system in an advanced phase of the merger, places itself close to non-interacting systems on the M$^{\rm diff}$-SFR relation. On the other hand, looking at the third panel which shows M$^{\rm diff}$ as a function of M$_\star$, we see that interacting systems have $\sim1.4\times$ more carbon mass in the diffuse halo than single galaxies. This result suggests that stripping mechanisms seem to be responsible for bringing metals into the CGM of the analyzed systems. This conclusion is further corroborated by the fact that the fraction of SF clumps and carbon mass present in the diffuse halo of non-interacting galaxies is comparable to that of interacting galaxies, as shown in Figure \ref{fig:starF}.

In summary, it seems that for the interacting systems analyzed in this work, stripping mechanisms are responsible not only for bringing carbon into larger scales but also for enhancing [CII] emission possibly via shocks (see also \citealt{Ginolfi+2020a}). 

The interacting systems that we have analyzed could be considered as high-$z$ analogs of circumgalactic stripped carbon and shock-induced [CII] emission in local studies \citep{Appleton+2013, VelusamyLanger2014}. These are qualitative conclusions, based on a comparison between the properties of the inner CGM found around single and major merging galaxies in our simulation, and need to be reinforced by performing a more quantitative analysis of the specific effects of all feedback processes at play, both mechanical and radiative, which is beyond the scope of the present work. 
  
\section{Summary and conclusions}
\label{sec:conclusion}
In this paper, we analyze a sample of merging galaxies observed by the ALMA-ALPINE survey at redshift 4.5 < $z$ < 5.1, to investigate the [CII] emission coming from the gas \textit{between} the galaxies. We complement the observational analysis with cosmological simulations, specifically the hydrodynamical code \texttt{dustyGadget}, looking for synthetic merging systems that helped us interpret the nature of such emission. Our main results can be summarized as follows:   

\begin{enumerate}
    \item We analyze major merging systems in the ALPINE survey and confirm the presence of an extended (> 20 kpc) [CII] halo in interacting galaxies. This extended halo is larger than the one observed in isolated galaxies as a consequence of the dynamic interaction between galaxies.
    \item We find that at least 25 \% of the total [CII] emission associated with these systems comes from the medium surrounding the merging system, thus between the galaxies. 
    \item We establish the presence of correlations, either strong or tentative, between the amount (fraction) of [CII] emission from the diffuse halo and integrated (relative) properties of galaxies.
    \item We find that an extended carbon-rich envelope is present in interacting systems selected from the \texttt{dustyGadget} simulation as well. In particular, extended emissions of around 10 kpc and larger than 20 kpc have been found in non-interacting and interacting systems respectively. Also, we find strong correlations between diffuse carbon in the cold gas phase - used as a proxy for [CII] emission - and the physical properties of galaxies, in particular, with their total carbon mass, SFR, and stellar mass, suggesting that different mechanisms reside behind [CII] emission.  
    \item Using \texttt{dustyGadget} we investigate the nature of the metal-enriched envelope, which, apart from outflow mechanisms, can be attributed to dynamical interaction between merging galaxies, that extract carbon-rich gas out of galaxies, and the presence of star-forming satellites, that enrich the inner CGM with newborn stellar populations. We argue that most of the [CII] emission observed in the ALPINE systems originates from gas stripping mechanisms in turbulent collisional environments (see discussion in Section \ref{sec:discussion}), in analogy with broad [CII] emission observations of tidal tails of shock-excited carbon in local groups \citep{Appleton+2013}. 
    
\end{enumerate}

Altogether our findings suggest that dynamical interactions and star-forming clumps at high-$z$ can be an efficient mechanism for extracting gas out of galaxies and enriching the CGM with chemically evolved material. Deeper and higher resolution ALMA data and state-of-the-art simulations with refined sub-grids models are necessary to study more in detail the key role of mergers in the baryon cycle of distant galaxies and to understand the specific role of outflows, tidal interactions, and star-forming satellites in the observed [CII] diffuse emission. Concerning the observations, the first results from the ALMA-CRISTAL survey (PI: Herrera-Camus) have already provided us with great details on the extended [CII] emission from high-$z$ galaxies (see for example \citealt{Solimano+2024, Posses+2024}) laying the ground for even more in-depth observations and analyses. Moreover, ALMA observations together with future sub-mm information from single-antenna telescopes (e.g. AtLAST; \citealt{Mroczkowski+2024,vanKampen+2024,Lee+2024}) would considerably improve our knowledge of the CGM and of the [CII] emission in the high-$z$ Universe.

\begin{acknowledgements}
      The authors would like to thank the anonymous referee for the useful suggestions which improved this article. This paper is based on data obtained with the ALMA Observatory, under Large Program 2017.1.00428.L. ALMA is a partnership of ESO (representing its member states), NSF (USA), and NINS (Japan), together with NRC (Canada), MOST and ASIAA (Taiwan), and KASI (Republic of Korea), in cooperation with the Republic of Chile. The Joint ALMA Observatory is operated by ESO, AUI/NRAO and NAOJ. CDC would like to thank the GESO group at the European Southern Observatory (ESO) for the useful discussions while preparing this manuscript. The simulated data underlying this article will be shared on reasonable request to the corresponding author. CDC acknowledged support from Sapienza University of Rome program "Bando per la mobilità individuale all'estero" (DR n. 1607 del 14 June 2021) during the visiting period (June-November 2022) at ESO Garching, Germany. LG and RS acknowledge support from the PRIN 2022 MUR project 2022CB3PJ3 - First Light And Galaxy aSsembly (FLAGS) funded by the European Union – Next Generation EU, and from the Amaldi Research Center funded by the MIUR program "Dipartimento di Eccellenza" (CUP:B81I18001170001). MR acknowledges support from the Narodowe Centrum Nauki (UMO-2020/38/E/ST9/00077) and support from the Foundation for Polish Science (FNP) under the program START 063.2023. We have benefited from the publicly available software CASA and CARTA and programming language \texttt{Python}, including the \texttt{numpy} (\url{https://numpy.org}), \texttt{matplotlib} (\url{https://matplotlib.org}), \texttt{scipy} (\url{https://scipy.org}) and \texttt{astropy} (\url{http://www.astropy.org}) packages.
\end{acknowledgements}

\bibliographystyle{aa}
\bibliography{mergingGalaxies}

\begin{appendix}
\section{[CII] velocity shifts}
\label{app:vel_shift}
\begin{figure*}[h!]
\centering
\includegraphics[scale = 0.35]{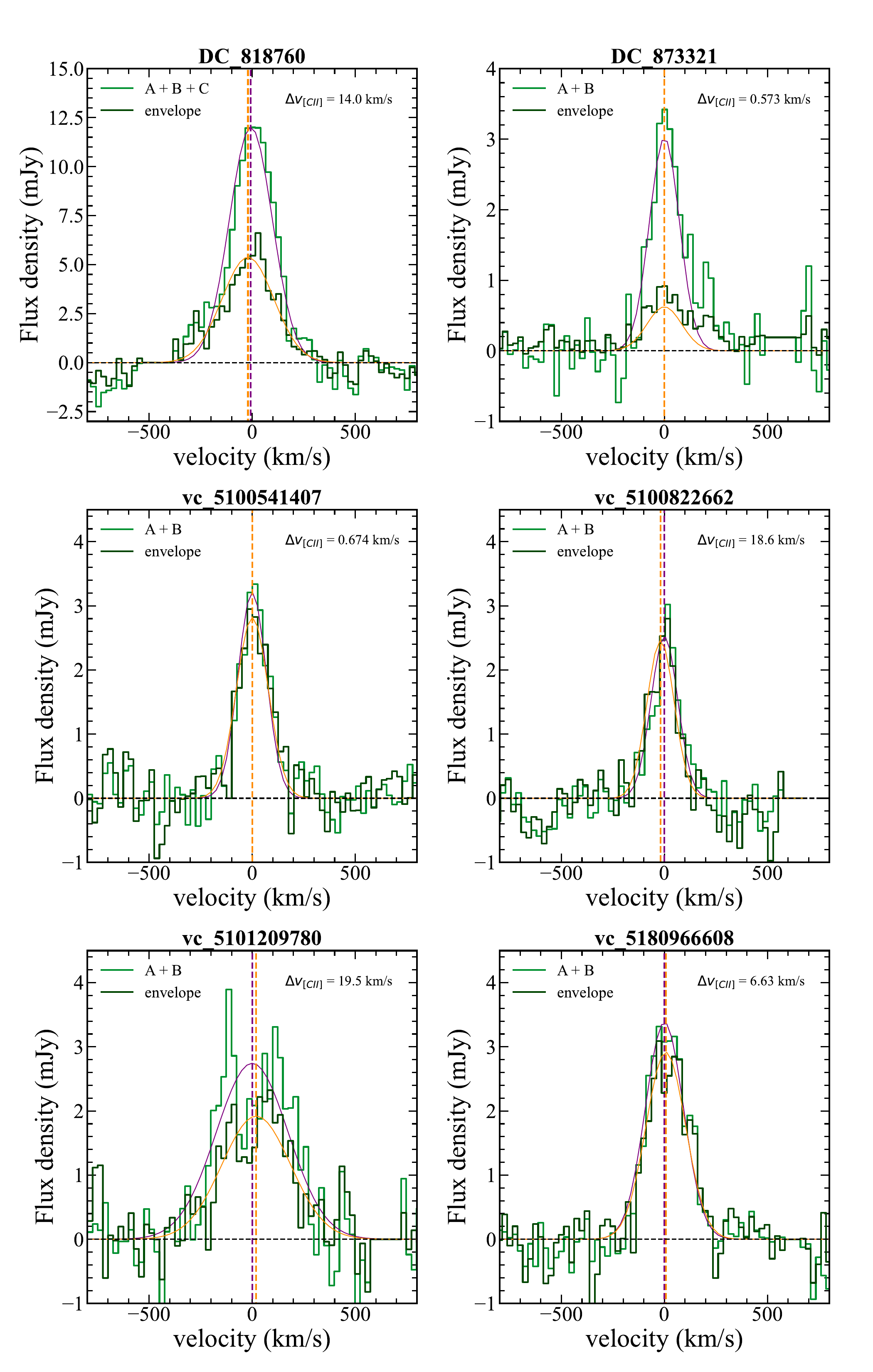}
\caption{The [CII] flux density (mJy) as a function of velocity (km/s) for our observational sample. The [CII] emission from the diffuse gas (i.e. envelope) is depicted in dark green, while in light green is the emission associated with the galaxies - assuming the FWHM-based aperture case. The emission associated with the envelope has been corrected for the 0.58 factor which takes into account that part of the emission can attributed to the tails of the PSF (see Section \ref{sec:diffEmission} for discussion). Gaussian fits for the emission associated with the diffuse gas component and galactic one are in orange and pink respectively. The dashed vertical lines show the central values of the Gaussian fits. The velocity shift ($\Delta v_{\rm [CII]}$) between these two components is displayed on the top right side of each panel.}
\label{fig:vel_shift}
\end{figure*}

In this section we test the presence, in any, of a velocity shift among the diffuse [CII] component and that associated with the merging galaxies. Figure \ref{fig:vel_shift} shows the [CII] flux density as a function of velocity for our sample: the diffuse component (after accounting for the possible effect of the PSF broadening, see discussion in Section \ref{sec:diffEmission}) is shown in dark green while that associated to galaxies in light green. We fit these emissions with a 1D-Gaussian function and look for any shift among its central values. The velocity shift ($\Delta v_{\rm [CII]}$) is reported in the top right corner of each panel. In general, we find $\Delta v_{\rm [CII]}$ $\lesssim 19$ km/s which suggests a small difference in the kinematics of the [CII] emission associated with the diffuse medium and that coming from the galaxies. This result is in line with what has been found previously for ALMA-ALPINE galaxies (see for example \citealt{Ginolfi+2020b} for merging systems and \citealt{Fujimoto+2020} for individual galaxies), and suggests that these two regions have a similar bulk motion.

\section{The merging system H6}
\label{app:A}
\begin{figure*}[h!]
\centering
\includegraphics[scale = 0.35]{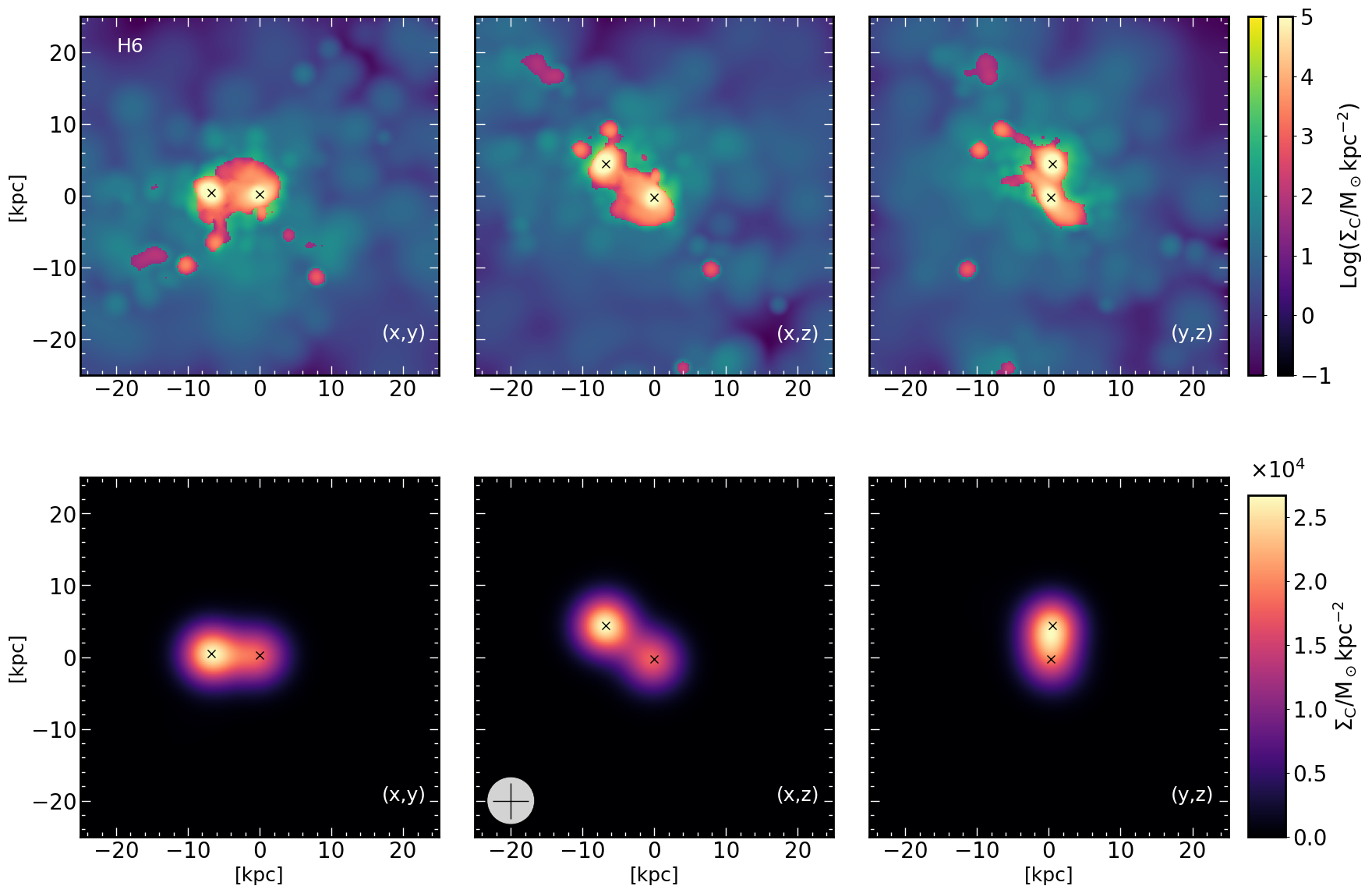}
\caption{The carbon surface density distribution in the synthetic merging system H6, each column is a projection. \textit{Top row:} The green color map shows the \textit{total} surface density distribution of carbon and overplotted in orange tones is the carbon distribution in the cold gas phase (T < $5 \times 10^4$ K). \textit{Bottom row:} Surface density distribution of carbon once we convolve the original map with the ALPINE-ALMA PSF. Black crosses indicate each merging galaxy and in the bottom left corner is a beam of $\sim 1 "$ as a reference. The x-projection in the bottom-right panel shows that it is difficult to discriminate between the two interacting galaxies.}
\label{fig:H6}
\end{figure*}
The synthetic merging system H6 has been considered as a peculiar system and excluded from the statistical analysis previously performed, as the two interacting galaxies are very close to each other (mean distance value: r$_m$ = 5.3 kpc, see Table \ref{tab:simGal}) resulting in a smaller fraction of carbon associated to the diffuse envelope. Indeed, it is difficult to distinguish the two merging galaxies, as shown by the x-projection in the smoothed case (see the bottom right panel in Figure \ref{fig:H6}). For this reason, we do not consider it as belonging to the "fiducial sample" of galaxies. In this system the interaction between galaxies is in such an advanced phase that the carbon mass belonging/associated to the galaxies and that of the diffuse envelope are mixed together, leading to a tricky analysis and interpretation. Moreover, the system appears differently depending on the chosen projection, resulting in large error bars (see Figures \ref{fig:diffuse_sim} and \ref{fig:fraction_sim}). This peculiar system, which can be considered as the synthetic counterpart of DC\_873321 (see Section \ref{sec:diffEmission}) needs to be treated and interpreted with care.  

\section{Carbon fraction versus average distance between galaxies}
\label{app:B}

\begin{figure*}[b]
    \begin{minipage}{.6\textwidth}
    \includegraphics[scale = 0.4]{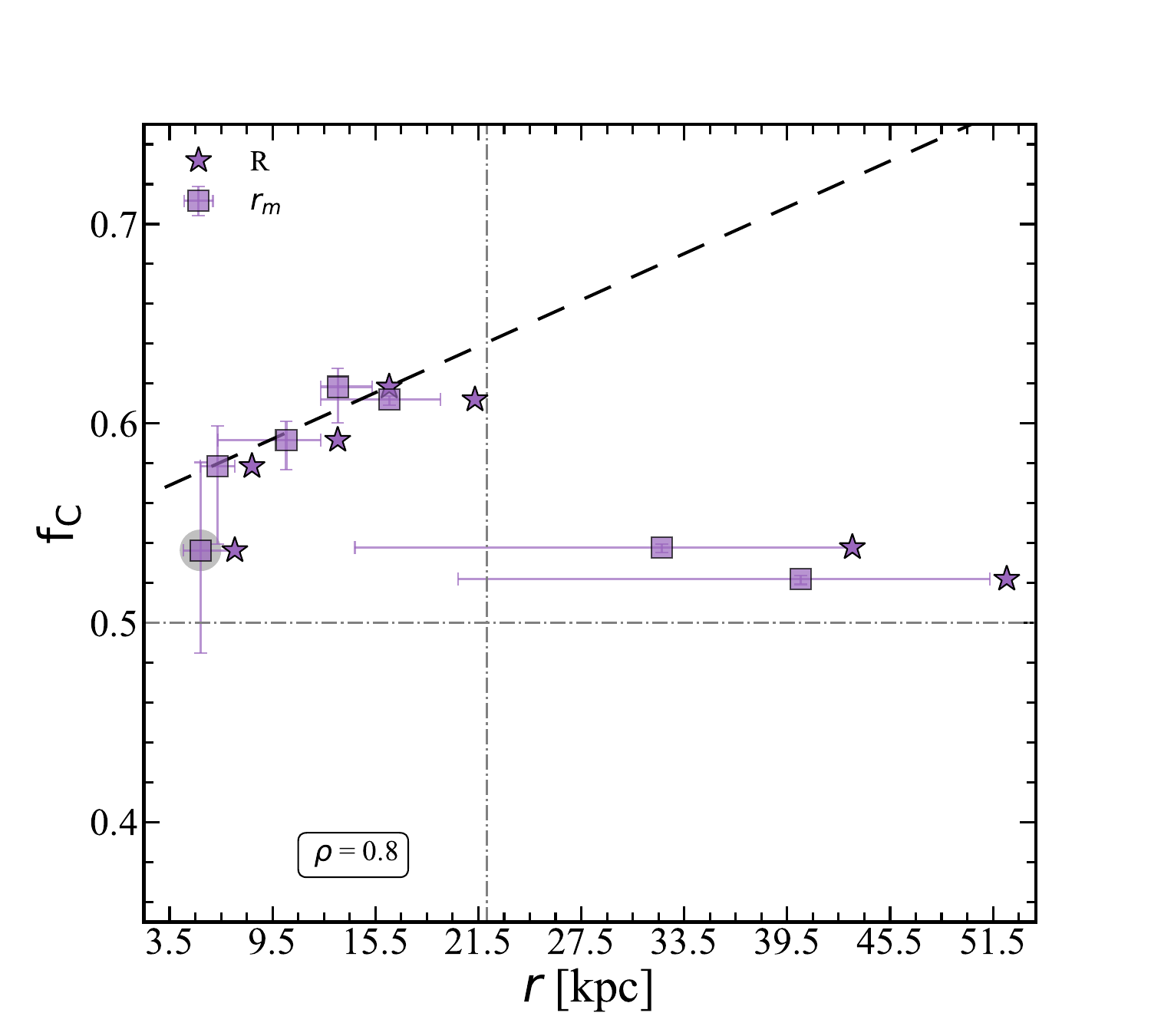}
    \end{minipage}
    \begin{minipage}{.4\textwidth}
    \caption{Fraction of carbon mass in the cold gas phase as a function of the mean distance (r$_m$, violet square) and physical distance (R, violet stars) between the galaxies. The horizontal and vertical gray dot-dashed lines show the f$_C$ = 0.5 value and $r = 21.5$ kpc, the maximum value for the x-axis in Figure \ref{fig:CGMfraction}, respectively.}
    \label{fig:FcR}
    \end{minipage}
\end{figure*}

In the left panel of Figure \ref{fig:fraction_sim} we show the correlation between the fraction of carbon in the cold gas phase that resides in the diffuse envelope and the mean distance between the merging galaxies ($r_m$, averaged between three projections). The relation between these two quantities leads to a correlation with $\rho = -0.5$ and a negative trend which seems to disagree with what has been found from the analysis of the observational sample ($\rho$ = 0.9). However, by further investigating this trend we find that the positive or negative correlation depends on the considered range of values for $r_m$. In fact, as shown in Figure \ref{fig:FcR}, if we only consider $r_m$ up to $\sim 16$ kpc (maximum value for the projected distance in the ALPINE sample), we recover the positive trend ($\rho = 0.8$) found in observations. This result highlights that the distance between the galaxies, thus the amount of time galaxies have been interacting, plays a fundamental role in the pollution of their inner CGM, suggesting that systems in their early phase of interaction (at larger distances) have less carbon-rich envelopes.

\end{appendix}
\end{document}